\documentclass{article}[12pt]
\usepackage {epsfig,subfigure,english,natbib}

\newcommand{\ppp}{\pi^{+} \pi^{-} \pi^{0}}
\newcommand{\mmg}{\mu^{+} \mu^{-}\gamma}
\newcommand{\eeg}{e^{+} e^{-}\gamma}
\newcommand{\noi}{\noindent}
\newcommand{\pip}{\ensuremath{\pi^+\,}}
\newcommand{\pim}{\ensuremath{\pi^-\,}}
\newcommand{\piz}{\ensuremath{\pi^0\,}}
\newcommand{\Ks}{\ensuremath{K_S \,}}
\newcommand{\Kl}{\ensuremath{K_L\,}}

\newcommand{\Eta}{\ensuremath{\eta\,}}
\newcommand{\etap}{\ensuremath{\eta'\,}}
\newcommand{\phot}{\ensuremath{\gamma\,}}
\newcommand{\tp}{\ensuremath{\vartheta_P \,}}

\newcommand{\cm}{\ensuremath{\,{\rm cm}}}

\newcommand{\s}{\ensuremath{\,{\rm s}}}

\newcommand{\pb}{\ensuremath{\,{\rm pb}}}
\newcommand{\pbinv}{\ensuremath{\,{\rm pb}^{-1}}}

\newcommand{\MeV}{\ensuremath{\,{\rm MeV}}}
\newcommand{\fikskl}{\ensuremath{\phi\rightarrow\Ks\Kl}}
\newcommand{\fietag}{\ensuremath{\phi\rightarrow\eta\gamma\;}}
\newcommand{\fietapg}{\ensuremath{\phi\rightarrow\eta'\gamma\;}}
\newcommand{\etapippimpiz}{\ensuremath{\eta\rightarrow\pip\pim\piz}}
\newcommand{\etapizpizpiz}{\ensuremath{\eta\rightarrow\piz\piz\piz}}
\newcommand{\etagg}{\ensuremath{\eta\rightarrow\gamma\gamma\;}}
\newcommand{\etappippimeta}{\ensuremath{\etap\rightarrow\pip\pim\eta}}
\newcommand{\etappizpizeta}{\ensuremath{\etap\rightarrow\piz\piz\eta}}

\title{
\begin{flushright}
\small Kloe Memo 00-215
\end{flushright}
\begin{flushright}
\small June 2000
\end{flushright}
KLOE first results on hadronic physics}
\date{ }
\author{The KLOE collaboration}

\begin{document}
\maketitle

\begin{abstract}

The KLOE detector \cite{bertolucci} at DA$\Phi$NE, the Frascati  
$\phi$-factory, has started
taking data in April 1999 and a total integrated luminosity of $2.4$ pb$^{-1}$
has been collected by the end of '99, corresponding to $\sim$8 millions $\phi$
decays.
With these data a preliminary measurement of $\phi$ radiative decays 
 $\phi \rightarrow \eta \gamma$, $\phi \rightarrow \pi^{0} \gamma$,
 $\phi \rightarrow \eta' \gamma$, $\phi \rightarrow \pi^{0}\pi^{0} \gamma$,
 $\phi \rightarrow \pi^{+}\pi^{-}  \gamma$, 
$\phi \rightarrow \eta\pi^{0} \gamma$ and of the hadronic decay
 $\phi \rightarrow \pi^{+}\pi^{-}\pi^{0}$ has been performed.
The energy spectrum of the radiated photon in case of the 
 $\pi^{0}\pi^{0}\gamma$, $\pi^{+}\pi^{-} \gamma$, $\eta\pi^{0}\gamma$ 
final states allows us to extract the information on the 
contribution of the direct decays 
 $\phi \rightarrow f_0 \gamma$, $\phi \rightarrow a_0 \gamma$. 
The measurement of BR( $\phi \rightarrow f_0 \gamma$), 
BR($\phi \rightarrow a_0 \gamma$) can help in understanding the nature of
$f_0(980)$ and $a_0(980)$ which is still under debate.
The value of BR($\phi \rightarrow \eta' \gamma$) can be related to
the gluonic content of the  $\eta'$(958) while  the ratio 
R=BR($\phi \rightarrow \eta' \gamma$)/BR($\phi \rightarrow \eta \gamma$)
can help in establishing the value of the  $\eta-\eta'$ mixing angle  $\theta_p$. 
\newline
Furthermore a high statistics analysis of the Dalitz plot in the 
 $\phi \rightarrow \pi^{+}\pi^{-}\pi^{0}$ decay allows us to extract a
possible contribution of the direct decay with respect to the dominant 
 $\rho\pi$ mode and to obtain a new measurement of the parameters of the 
$\rho$ line shape, including the $\rho^{0}-\rho^{\pm}$ mass difference.

\end{abstract}

\vspace*{6truecm}
\begin{center}
{\small  Contributed paper N.220 to the XXX International Conference on High
  Energy Physics, \newline Osaka 27 jul - 2 aug 2000.
}
\end{center}

\newpage

\def\ifm#1{\relax\ifmmode#1\else$#1$\fi}
\def\eps{\ifm{\epsilon}} \def\epm{\ifm{e^+e^-}}
\def\rep{\ifm{\Re(\eps'/\eps)}}  \def\imp{\ifm{\Im(\eps'/\eps)}}  
\def\DAF{DA$\Phi$NE}  \def\sig{\ifm{\sigma}}
\def\gam{\ifm{\gamma}} \def\to{\ifm{\rightarrow}}
\def\pip{\ifm{\pi^+}} \def\pim{\ifm{\pi^-}}
\def\po{\ifm{\pi^0}} 
\def\pic{\ifm{\pi^+\pi^-}} \def\pio{\ifm{\pi^0\pi^0}} 
\def\ks{\ifm{K_S}} \def\kl{\ifm{K_L}} \def\kls{\ifm{K_{L,\,S}}} 
\def\ksl{\ifm{K_S,\ K_L}} \def\ko{\ifm{K^0}}
\def\K{\ifm{K}} \def\LK{\ifm{L_K}}
\def\Kb{\ifm{\rlap{\kern.3em\raise1.9ex\hbox to.6em{\hrulefill}} K}}
\def\ab{\ifm{\sim}}  \def\x{\ifm{\times}}
\def\ff{$\phi$--factory}
\def\sta#1{\ifm{|\,#1\,\rangle}} 
\def\amp#1,#2,{\ifm{\langle#1|#2\rangle}}
\def\kob{\ifm{\Kb\vphantom{K}^0}}
\def\f{\ifm{\phi}}   \def\pb{{\bf p}}
\def\L{\ifm{{\cal L}}}  \def\R{\ifm{{\cal R}}}
\def\up#1{$^{#1}$}  \def\dn#1{$_{#1}$}
\def\etal{{\it et al.}}
\def\BR{{\rm BR}}
\def\radl{\ifm{X_0}}
\def\deg{\ifm{^\circ}} 
\def\th{\ifm{\theta}}
\def\To{\ifm{\Rightarrow}}
\def\ot{\ifm{\leftarrow}}
\def\fo{\ifm{f_0}} \def\epe{\ifm{\eps'/\eps}}
\def\pbrn{ {\rm pb}}  \def\cm{ {\rm cm}}
\def\mub{\ifm{\mu{\rm b}}} \def\s{ {\rm s}}
\def\RR{\ifm{{\cal R}^\pm/{\cal R}^0}}
\def\dt{ \ifm{{\rm d}t} } \def\dy{ {\rm d}y } \def\pbrn{ {\rm pb}}
\def\kp{\ifm{K^+}} \def\km{\ifm{K^-}}
\def\kkb{\ifm{\ko\kob}} 
\def\epe{\ifm{\eps'/\eps}}
\def\ppc{\ifm{\pi^+\pi^-}}
\def\ppo{\ifm{\pi^0\pi^0}}
\def\pppco{\ifm{\pi^+\pi^-\pi^0}}
\def\pppo{\ifm{\pi^0\pi^0\pi^0}}
\def\vare{\ifm{\varepsilon}}
\def\etap{\ifm{\eta'}}

\def\pt#1,#2,{#1\x10\up{#2}}

\def\B{Bari}
\def\b{\rlap{\kern.2ex\up a}}
\def\O{IHEP}
\def\o{\rlap{\kern.2ex\up b}}
\def\Fr{Frascati}
\def\fr{\rlap{\kern.2ex\up c}}
\def\Ka{Karlsruhe}
\def\ka{\rlap{\kern.2ex\up d}}
\def\Le{Lecce}
\def\le{\rlap{\kern.2ex\up e}}
\def\Mo{Moscow}
\def\mo{\rlap{\kern.2ex\up f}}
\def\N{Napoli}
\def\n{\rlap{\kern.2ex\up g}}
\def\BE{Beer-Sheva}
\def\be{\rlap{\kern.2ex\up h}}
\def\co{\rlap{\kern.2ex\up i}}
\def\Pi{Pisa}
\def\pI{\rlap{\kern.2ex\up j}}
\def\Ra{Roma I}
\def\ra{\rlap{\kern.2ex\up k}}
\def\en{\rlap{\kern.2ex\up l}}
\def\Rb{Roma II}
\def\rb{\rlap{\kern.2ex\up m}$\,$}
\def\Rc{Roma III}
\def\rc{\rlap{\kern.2ex\up n}}
\def\su{\rlap{\kern.2ex\up o}}
\def\T{Trieste/Udine}
\def\t{\rlap{\kern.2ex\up q}}
\def\V{Virginia}
\def\v{\rlap{\kern.2ex\up r}}
\def\Z{Associate member}
\def\hsa{ \ }

%\author{ 
%{\baselineskip 12pt
%\parskip=0pt
%\parindent=0pt}
\normalsize\noindent
M.~Adinolfi\rb,\hsa 
A.~Aloisio\n,\hsa
F.~Ambrosino\n,\hsa
A.~Andryakov\mo,\hsa
A.~Antonelli\fr,\hsa %\\[-1.5mm] \normalsize
M.~Antonelli\fr,\hsa 
F.~Anulli\fr,\hsa
C.~Bacci\rc,\hsa
A.~Bankamp\ka,\hsa
G.~Barbiellini\t,\hsa
% ...
%}  %%%%

%\maketitle
\vglue 2mm

\def\aff#1{Dipartimento di Fisica dell'Universit\`a e Sezione INFN, #1, Italy.}

%{\baselineskip=12pt
%\parskip=0pt
%\parindent=0pt
\def\hsb{\hskip 2.8mm}

\leftline{\b\hsb \aff{\B}}
\leftline{\o\hsb Institute of High Energy Physics of Academica Sinica, 
Beijing, China.}
\leftline{\fr\hsb  Laboratori Nazionali di Frascati dell'INFN, Frascati, Italy.}
\leftline{\ka\hsb  Institut f\"ur Experimentelle Kernphysik, Universit\"at \Ka,
Germany.}
\leftline{\le\hsb \aff{\Le}}
\leftline{\mo\hsb Institute for Theoretical and Experimental Physics, Moscow,
Russia.}
\leftline{\n\hsb Dipartimento di Scienze Fisiche dell'Universit\`a e 
Sezione INFN, \N, Italy.}
\leftline{\be\hsb Physics Department, Ben-Gurion University of the Negev,
Israel.}
\leftline{\co\hsb Physics Department, Columbia University, New York, USA.}
\leftline{\pI\hsb \aff{\Pi}}
\leftline{\ra\hsb \aff{\Ra}}
\leftline{\en\hsb Dipartimento di Energetica dell'Universit\`a, Roma I, Italy.}
\leftline{\rb\hsb \aff{\Rb}}
\leftline{\rc\hsb \aff{\Rc}}
\leftline{\su\hsb Physics Department, State University of New York 
at Stony Brook, USA.}
\leftline{\t\hsb \aff{\T}}
\leftline{\v\hsb Physics Department, University of Virginia, USA.}
\leftline{\rlap{\kern.2ex\up*}\hsb\Z}    
%  }  

\newpage

\section{Introduction}

\subsection{Radiative decays}

\subsubsection{The scalar sector: $\phi\rightarrow f_0\gamma$, 
$\phi\rightarrow a_0\gamma$ }

The lightest scalar mesons, with masses below 1 GeV, have defied
classification for nearly half a century. The narrow states $f_0(980)$ and
$a_0(980)$ do not conform to standard $q\overline{q}$ quark model
expectations. 
The biggest departures from theoretical predictions based on the
$q\overline{q}$ model are in the total width (predicted
$\Gamma\sim$500 MeV, observed $\Gamma\sim$50 MeV) and in the $\gamma
\gamma$ coupling (predicted $\sim$4.5 keV for $f_0$, $\sim$1.5 keV for
$a_0$, observed $\leq$0.6 keV for $f_0$, $\sim$0.2 keV for $a_0$).
Two alternative hypotheses 
\cite{close} are presently under discussion for their nature: they could
be a four quark state $q\overline{q}q\overline{q}$ (R$\sim$1 fm)
\cite{jaffe}
or a $K\overline{K}$ molecule (R$\sim$1.7 fm)
\cite{isgur}.
Different BR's are expected depending on their nature, as can be seen
in Table \ref{tabada} in the case of the $f_0$.

\begin{table}[h]
\begin{center}
\begin{tabular}{|c|c|} \hline
Model & BR($\phi \rightarrow f_0 \gamma$) \\ \hline
$s\overline{s} (^{3}P_0$)  &  $\sim 10^{-5}$\\ \hline
($u\overline{u}+d\overline{d})/\sqrt{2} (^{3}P_0$)  &  $\leq 10^{-6}$\\ \hline
$q\overline{q}q\overline{q}$  &  $\sim 10^{-4}$\\ \hline
$K\overline{K}$ molecule  &  $10^{-4}\div 10^{-5}$\\ \hline
\end{tabular}
\caption{Theoretical predictions for $\phi \rightarrow f_0 \gamma$.}
\label{tabada}
\end{center}
\end{table}

Recent studies using lattice QCD \cite{jaffe2000} suggest that  
$q\overline{q}q\overline{q}$ states occurs generically near meson-meson 
thresholds. The recent observation of  $\phi\rightarrow f_0\gamma$, 
$\phi\rightarrow a_0\gamma$ with B.R. $\sim 10^{-4}$\cite{cmd2_f0,snd_scal}
seems to be 
in favour of the $q\overline{q}q\overline{q}$ scenario.
The observation from Crystal Barrel of an isoscalar
$f_0$(1365) \cite{cbarrel1} and an isovector $a_0$(1450) \cite{cbarrel2}
as members of the $^{3}P_0$ nonet
allows us to search for an explanation for the $f_0$(980), $a_0$(980) outside
the $q\overline{q}$ model.

\subsubsection{The pseudoscalar sector: $\phi\rightarrow\eta\gamma$, 
$\phi\rightarrow\eta'\gamma$ }

The reason for studying $\phi$ radiative decays in $\eta$ and $\eta$' is
twofold since the measurement of BR($\phi\rightarrow\eta'\gamma$) can help in
defining the gluonic content of the $\eta'$ while the measurement of the
ratio R=BR($\phi\rightarrow\eta'\gamma$)/  
BR($\phi\rightarrow\eta\gamma$) can help in establishing the value of the 
$\eta-\eta'$ mixing angle $\theta_P$.\newline
The presence of gluon admixture in the $\eta'$ wavefunction is a
longstanding problem that could be solved by an accurate measurement of 
BR($\phi\rightarrow\eta'\gamma$): theoretical predictions range from
as low as $10^{-6}$ in models with gluonium admixture \cite{deshp}
or with 
strong QCD violations \cite{benay1} to  $10^{-4}$ in different realizations of
the quark model \cite{q1,q2,q3,q4}. The recent measurement of
BR($\phi\rightarrow\eta'
\gamma$) by the CMD-2\cite{cmd2_etap} and SND\cite{snd_etap} collaboration
at VEPP-2M seems to exclude a gluonium admixture.  

The value of $\theta_P$ has been discussed many times in the last thirty
years \cite{m1,m2,m3,m4}: the quadratic Gell Mann Okubo mass formula gives  $\theta_P \sim
-10^{o}$ while experimental data give $\theta_P$  in the range from $-14^{o} to
-20^{o}$.
A recent analysis \cite{BramonEscribano}, based mainly on decays
$J/\psi\rightarrow VP$, give  $\theta_P$=-16.9 $\pm$ 1.7. 
A crucial test, originally proposed by Rosner \cite{q3} is the measurement
of the ratio  R=BR($\phi\rightarrow\eta'\gamma$)/ 
BR($\phi\rightarrow\eta\gamma$). This ratio predicts $7.6\times10^{-3}$ for  
$\theta_P \sim-20^{o}$ and $6.2\times10^{-3}$ for $\theta_P \sim-16.9^{o}$.

\subsection{$\phi\rightarrow\pi^+\pi^-\pi^0$}

About $15\%$ of the $\phi$ decay into $\pi^+\pi^-\pi^0$. These final
states can be due to three different mechanisms (see fig.\ref{Diagrams}):
\begin{enumerate}
\item{$\phi\rightarrow\rho\pi$ decay where $\rho\pi$ include all the three
possible charge states (namely $\rho^+\pi^-$, $\rho^0\pi^0$ and $\rho^-\pi^+$)
with the same isospin weights;}
\item{$\phi\rightarrow\pi^+\pi^-\pi^0$ direct decay;}
\item{$e^+e^-\rightarrow \omega\pi^0$, with $\omega\rightarrow\pi^+\pi^-$.}
\end{enumerate}

\begin{figure}[h]
\begin{tabular}{ccc}
 \epsfig{file=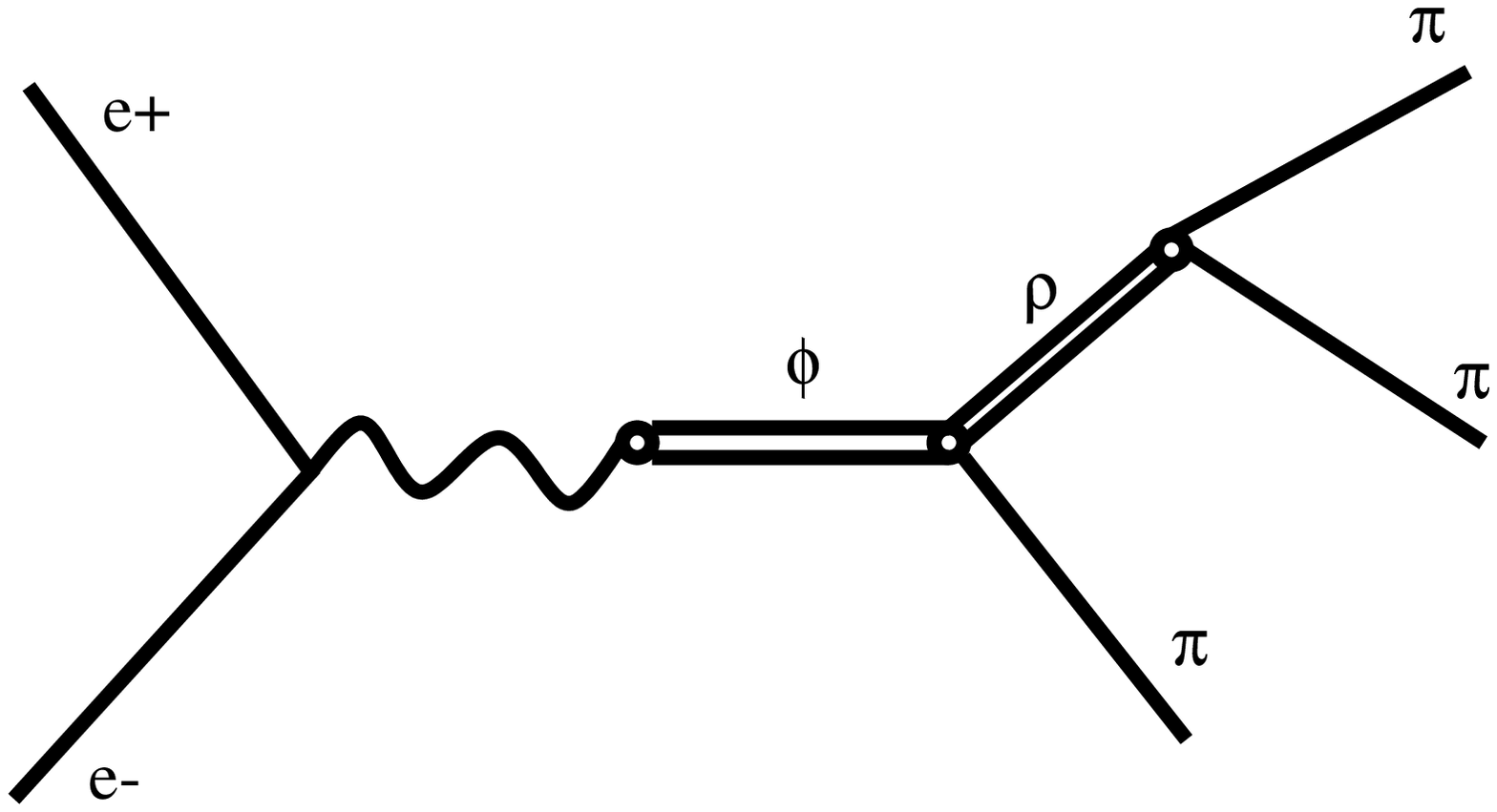,width=0.3\textwidth} & 
 \epsfig{file=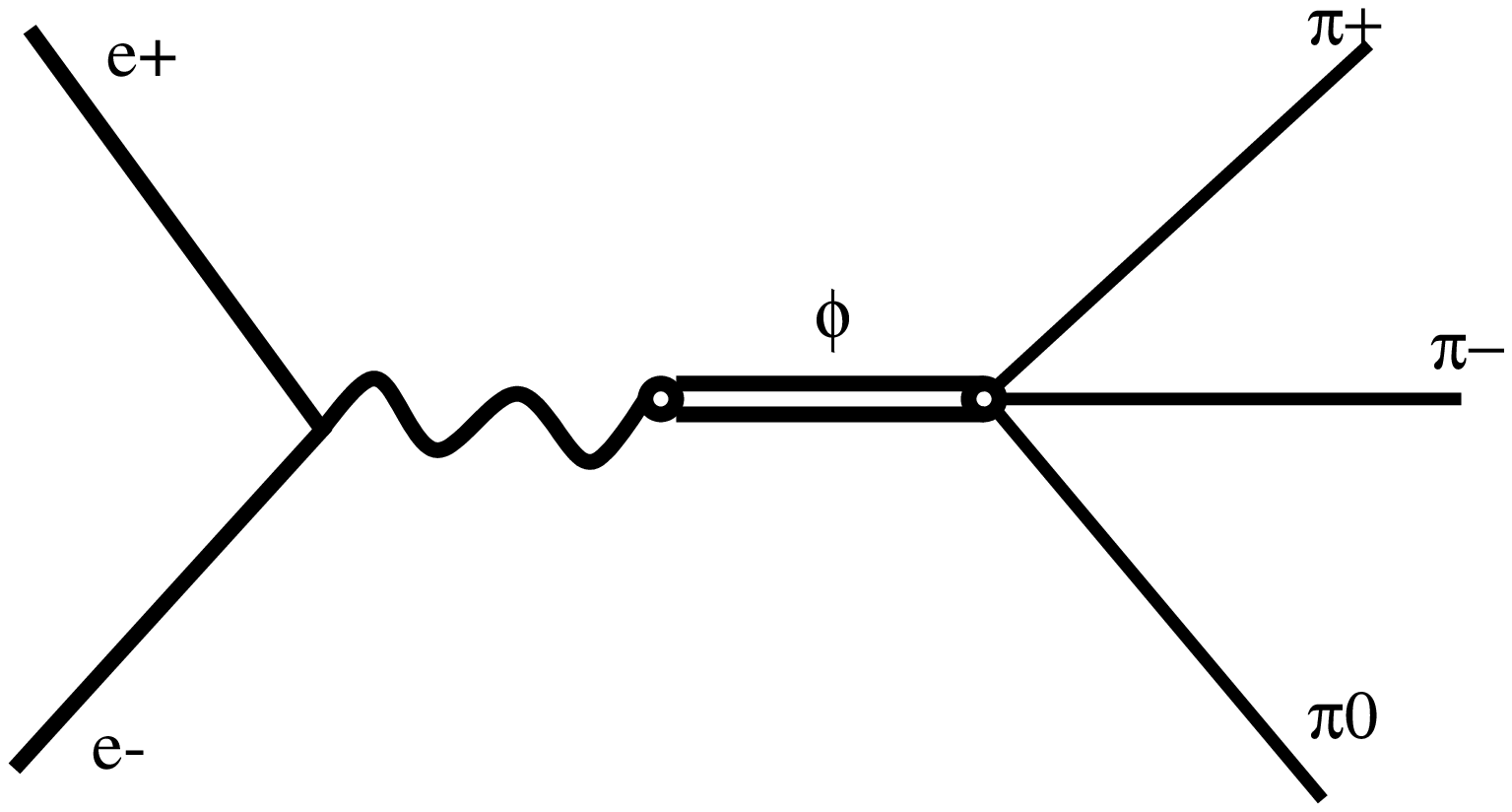,width=0.3\textwidth} &
\epsfig{file=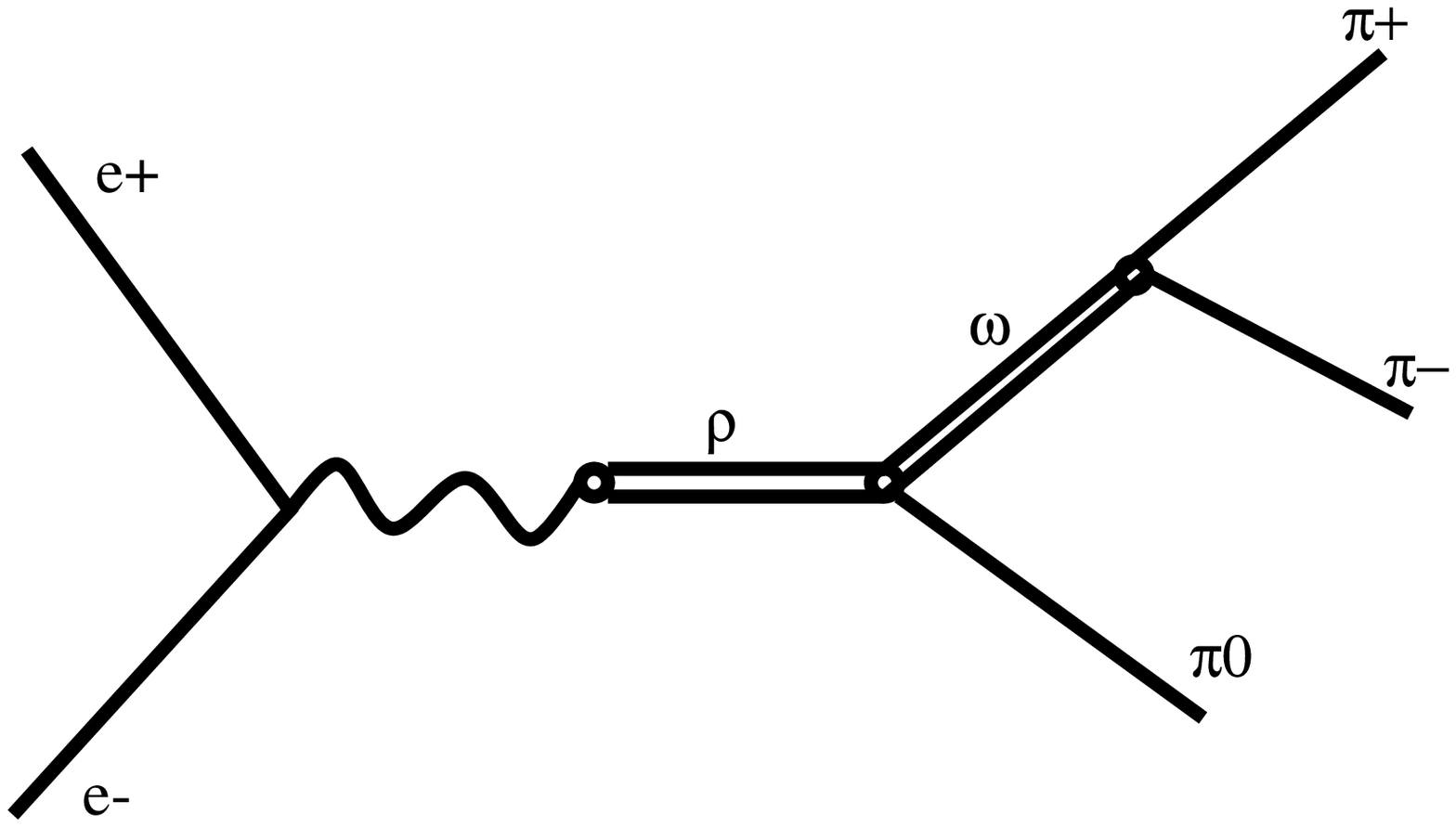,width=0.3\textwidth} \\
\end{tabular}
\caption{Feymnan diagrams contributing to the $\pi^+\pi^-\pi^0$ final state.}
\label{Diagrams}
\end{figure}
        
The fit of the Dalitz-plot distribution of this three-body decay allows us
to discriminate between the three contributions. In particular it is possible
to observe the direct contribution (number 2) as predicted by several
theoretical models, that has never been observed in
previous experiments. The only up to now published analysis of this kind
\cite{Novosippp} finds a Dalitz-plot fully dominated by the $\rho\pi$
contribution and only a limit of the direct decay to be below $\sim 10\%$.

Furthermore a precision measurement of the $\rho$ line shape parameters is
also possible. In particular the same amount of events corresponding to the
three charge states of the $\rho$, allows us to make comparisons between $\rho^+$
and $\rho^-$ (CPT test) and between charged $\rho$ and $\rho^0$. A mass or
width difference between $\rho^{\pm}$ and $\rho^0$ is a signature of isospin
violation as observed in other meson and baryon isospin multiplets.

\section{$\phi$ radiative decays}

\subsection{Selection criteria for radiative decays}
\label{kinefit}

Some steps of the analysis and some definitions are very similar for most
of the processes studied in this paper.

All the processes under study are characterized by the presence of prompt
photons, \rm{i.e.} photons coming from the the I.P.
These photons are detected as clusters in the calorimeter 
\footnote{The efficiency for photon detection is $\sim 85\%$ at 20 MeV and 
$>98\%$ for energies above 50 MeV.}
that obey the
relation $t-r/c=0$, where $t$ is the arrival time on the calorimeter, $r$
is the distance of the cluster from the I.P., and $c$ is the speed of light.
We define a photon to be ``prompt'' if $|t-r/c|<5\sigma_t$, where we use as
time resolution of the calorimeter the parameterization
$\sigma_t=110\hbox{ ps}/\sqrt{E\hbox{(GeV)}}$.
This $t-r/c$ interval is often referred to as ``time window'' in the
following sections.
An acceptance angular region corresponding to the polar angle interval 21$^o\div$ 159$^o$
is defined for the prompt photons, in order to exclude the blind region
around the beam-pipe.  

Most of the analyses described in this paper make use of a constrained fit
ensuring kinematic closure of the events.
The free parameters of the fit are: the three coordinates (x, y, z) of the
impact point on the calorimeter, the energy, and the time of flight for
each photon coming from the I.P., the track curvature and the two angles
$\phi$ and $\theta$ for each charged pion also coming from the I.P., the
two energies of the beams, and the three coordinates of the position of the
I.P.
The analysis procedure adopted is the following: 
\begin{enumerate}
        \item events with the appropriate number of prompt photons and charged tracks
              are selected from the ``radiative stream''; 
        \item the kinematic fit is applied on these events a first
              time with the constraints of the total energy and momentum
              conservation and satisfying $t-r/c=0$ 
              for each prompt photon;
        \item other selection criteria are applied to separate the
              signal from background;
        \item the kinematic fit is applied a second
              time on the surviving events with the same constraints as before plus other ones
              imposing the invariant masses of the particles present in the
              intermediate states ($\pi^0$'s, $\eta$'s etc.).
\end{enumerate}

\subsection{Luminosity measurement}

\begin{figure}[h]
        \begin{center}
        \epsfig{file=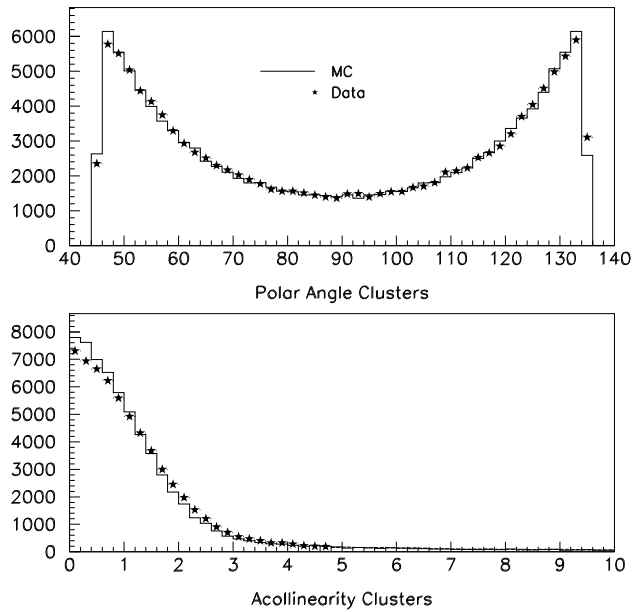,height=.35\textheight}
        \caption{Comparison between data and MC of polar angle and
acollinearity for Bhabha events.}
        \label{bhabha}
        \end{center}
\end{figure}

The integrated luminosity has been measured with large
angle Bhabha scattering events using only the calorimeter information. 
The measurement is described in detail in ref.\cite{lumi}. 
The absolute error on the luminosity measurement can be estimated in 
$\leq 3\%$.
Fig.\ref{bhabha} shows a comparison between data and Monte Carlo
for polar angle and acollinearity of large angle Bhabha
electrons and positrons, that are the relevant quantities for the
evaluation of the luminosity.
The agreement between the distributions is good. \newline

\subsection{$\phi\rightarrow\pi^0\pi^0\gamma$}

The $\pi^0\pi^0\gamma$ final state allows us to investigate the $f_0\gamma$
intermediate state.
The analysis scheme\cite{simona} has been developed using Monte Carlo
events, simulating  
both the signal and the main background channels with the S/B ratio
listed in  
Tab.\ \ref{Tab:Eff_fon}.
Photons coming from $\pi^0$'s have a flat 
energy distribution with $E_\gamma<500$ MeV while the radiative $\gamma$ is 
peaked at 50 MeV. The background spectrum covers the same energy range.
The kinematic fit, applied to the 5 photon final state, is relevant in
assigning the radiative photon and allows a partial rejection of the
background.  
After applying a first fit with constraints on quadri--momentum and time of 
flight, the best photons' combination producing two $\gamma\gamma$ pairs with 
$\pi^0$ mass and $M_{\pi^0\pi^0}>700$ MeV (the expected $f_0$ mass region)
is selected. A second fit, imposing further constraints on $\pi^0$'s mass on 
the assigned $\gamma\gamma$ pairs, is then performed without any assumption 
on $f_0$ mass and width.
This procedure correctly identifies the radiative photon in 92\% of the well
reconstructed $f_0\gamma$ events.

After the whole fit procedure, the resulting background rejection is still
not enough, especially for $\omega\pi^0$ (${\rm S/B}\sim 1$ after the fit).
A good variable which helps in identifying the $e^+e^-\to\omega\pi^0$ process 
is the angle $\psi$ between the primary photon and the pion's flight direction 
in the $\pi^0\pi^0$ rest frame. Because of the different spin between $f_0$ 
and $\omega$ ($J=0,\,1$ respectively), the $\cos\psi$ distribution is flat in 
the first case while is peaked at 0.5 in the second one 
(Fig.\ \ref{Fig:CosPsi}.left). 
Therefore the final kinematic fit is also performed with a different 
photon's assignment, requiring the best combination of four $\gamma$'s
into pions, which gives $M_{\pi^0\gamma}$ in agreement with the $\omega$ mass.
Using these criteria, the photon assignment for non-$\omega\pi^0$ events 
is not correct and the resulting distribution is peaked at high $\cos\psi$ 
values for $f_0\gamma$ and flat for the rest of the background
(Fig.\ \ref{Fig:CosPsi}.left).
$f_0\gamma$ candidates are then selected requiring $\cos\psi>0.8$ 
($f_0\gamma_{\rm cut}$) while the $\omega\pi^0$ selection requires
$0.4<\cos\psi<0.8$ ($\omega\pi^0_{\rm cut}$).

In Fig.\ \ref{Fig:CosPsi}.right, the $M_{\pi^0\pi^0}$ distribution in the 
$f_0\gamma$ fit hypothesis is shown for the signal before and after applying 
$f_0\gamma_{\rm cut}$. 
It is remarkable that this cut is very efficient and does not significantly
modify the $M_{\pi^0\pi^0}$ shape. 
In Tab.\ \ref{Tab:Eff_fon} the efficiencies for the various analysis steps 
as obtained from Monte Carlo (MC) are reported (cl: trigger, background
rejection and event classification;
\footnote{In the case of the $f_0\gamma$ signal the values of the
contributions to  $\varepsilon_{\rm cl}$ are the following:
 97.9\% from trigger, 90.2\% from background rejection, 86.7\% from event 
classification.}
sel+$\chi^2$: acceptance, time window and $\chi^2$ 
cuts; cut: $f_0\gamma_{\rm cut}$)
together with the signal/background ratios before and after the analysis.
Results for the $\omega\pi^0$ analysis are listed in Tab.\ \ref{Tab:Eff_wp}.
\footnote{In the case of the $\omega\pi^0$ signal the values of the
contributions to  $\varepsilon_{\rm cl}$ are the following:
 98.2\% from trigger, 89.1\% from background rejection, 90.7\% from event 
classification.}
\begin{figure}
\begin{tabular}{cc}
 \epsfig{file=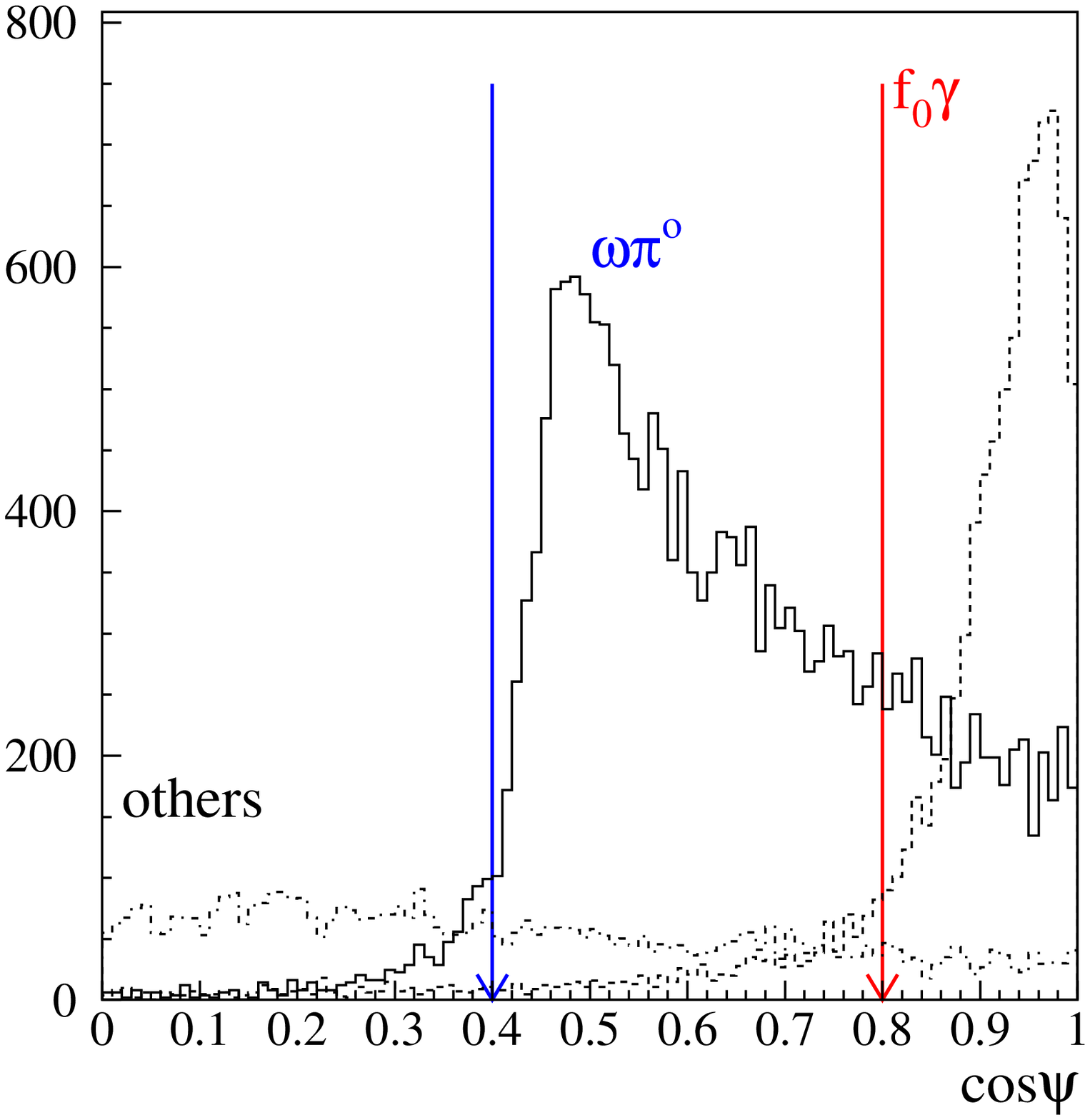,
   width=0.5\textwidth}                       &
 \epsfig{file=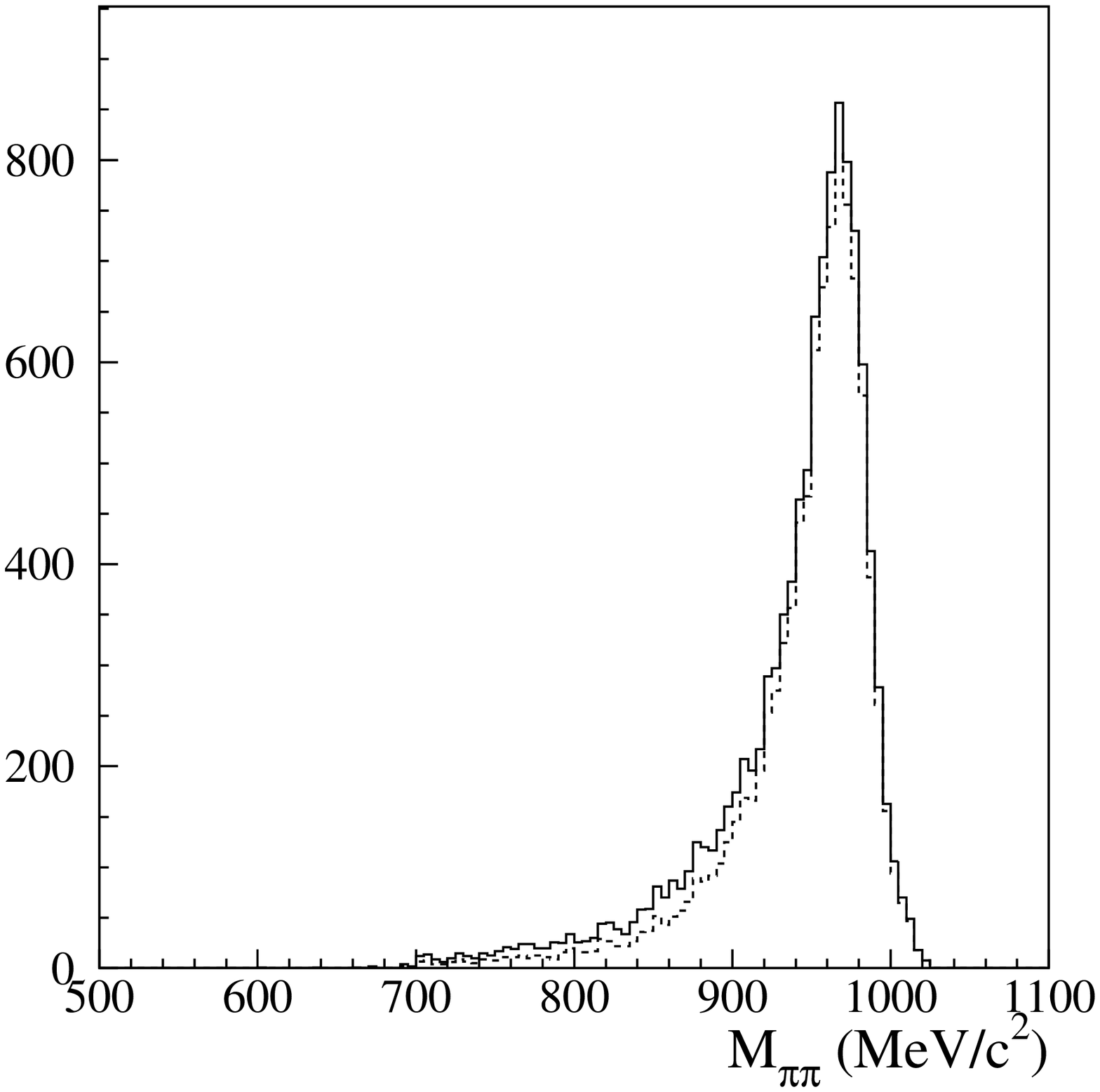,
   width=0.5\textwidth,height=0.5\textwidth}  \\
\end{tabular}
\caption{Monte Carlo distributions showing the effect of the cut on $\cos\psi$
         Left: angle between the primary photon and the pion's flight 
         direction in the $\pi^0\pi^0$ rest frame for $\omega\pi^0$ (solid), 
         $f_0\gamma$ (dashed) and the other background (dot--dashed). The
         fit is 
         performed in the $\omega\pi^0$  hypothesis.
         Right: $\pi^0\pi^0$ invariant mass using constrained variables
         after fitting in $f_0\gamma$ hypothesis, before (solid) and after 
         (dashed) the $f_0\gamma_{\rm cut}$ cut.}
\label{Fig:CosPsi}
\end{figure}
\begin{table}
\caption{Analysis efficiencies for $\pi^0\pi^0\gamma$ decay and related 
  background.}
\begin{center}
\begin{tabular}{|c|c|c|c|c|c|c|}\hline
Decay channel  &  S/B  &
 $\varepsilon_{\rm cl}$   &  $\varepsilon_{\rm sel+\chi^2}$ &
 $\varepsilon_{\rm cut}$  &  $\varepsilon_{\rm tot}$ &  Final S/B     \\ \hline
Signal           &  ---   & 76.6\% & 58.4\% & 88.6\% & 39.6\% &  ---  \\
$\omega\pi^0$    &  0.52  & 79.4\% & 30.9\% & 25.8\% &  6.3\% &  3.3  \\
$\rho\pi^0$      &  3.0   & 75.0\% & 22.8\% & 25.9\% &  4.4\% & 27.0  \\
$a_0\gamma$      &  3.1   & 68.3\% & 15.5\% & 26.8\% &  2.8\% & 43.8  \\
$\eta\gamma$     &  0.02  & 75.0\% & 0.3\%  & 45.9\% & $9\times 10^{-4}$ 
                                                              & 8.8  \\ \hline
\end{tabular}
\end{center}
\label{Tab:Eff_fon}
\end{table}
\begin{table}
\caption{Analysis efficiencies for $e^+e^-\to\omega\pi^0\to\pi^0\pi^0\gamma$ 
  decay and related background.}
\begin{center}
\begin{tabular}{|c|c|c|c|c|c|c|}\hline
Decay channel  &  S/B  &  
 $\varepsilon_{\rm cl}$   &  $\varepsilon_{\rm sel+\chi^2}$ &
 $\varepsilon_{\rm cut}$  &  $\varepsilon_{\rm tot}$ &  Final S/B   \\ \hline
Signal           & ---  & 79.4\% & 59.2\% & 75.0\% & 35.2\% &  ---  \\
$f_0\gamma$      & 1.9  & 76.6\% & 48.3\% & 11.4\% &  4.2\% & 15.9  \\
$\rho\pi^0$      & 6.0  & 75.0\% & 42.9\% & 40.5\% & 13.0\% & 16.2  \\
$a_0\gamma$      & 6.0  & 68.3\% & 17.2\% & 23.9\% &  2.8\% & 75.4  \\
$\eta\gamma$     & 0.04 & 75.0\% & 0.5\% & 43.1\% & $1.7\times 10^{-3}$ 
                                                            & 8.3  \\\hline
\end{tabular}
\end{center}
\label{Tab:Eff_wp}
\end{table}

The same analysis is performed on the 1.84 pb$^{-1}$ collected on 
December '99\cite{simona}. Out of the 51666 events with at least 5 neutral clusters,
37678 are in the acceptance angular region and are reduced to 1815 after
applying the time window requirement. Performing the $\omega\pi^0$ fit, 980 
events have a good $\chi^2$. The distribution of the $\cos\psi$ variable
for this  
sample is shown in Fig.\ \ref{Fig:Data_wp}.left together with the expected 
one from Monte Carlo.
Because of the excellent agreement, the $\omega\pi^0_{\rm cut}$ is performed.
A nice peak at the $\omega$ mass appears in the $M_{\pi^0\gamma}$ distribution 
of the surviving 529 events (Fig.\ \ref{Fig:Data_wp}.right).
Background evaluation from Monte Carlo gives a final signal counting of 
$436\pm 25\ ({\rm stat})$.
Assuming a global systematic error of 8\%%
\footnote{The preliminary estimate of the systematic error has the following
contributions: 5\% from classification, 2.5\% from clustering (50\% of the 
events with at least 1 cluster not correctly reconstructed), 4\% from fit 
(50\% of the wrong MC 5 photon's assignment) and 3\% from luminosity.}
and correcting for luminosity and analysis efficiency we quote a cross section 

\begin{equation}
\sigma(e^+e^-\to\omega\pi^0\to\pi^0\pi^0\gamma) = 
   (\,0.67\pm 0.04\,({\rm stat.})\pm 0.05\,({\rm syst.})\,)\ {\rm nb}
\label{Eqn:Sigma_wp}
\end{equation}

\noindent
in good agreement with the SND measurement\cite{OmegaPio_SND}.

\begin{figure}
\begin{center}
 \epsfig{file=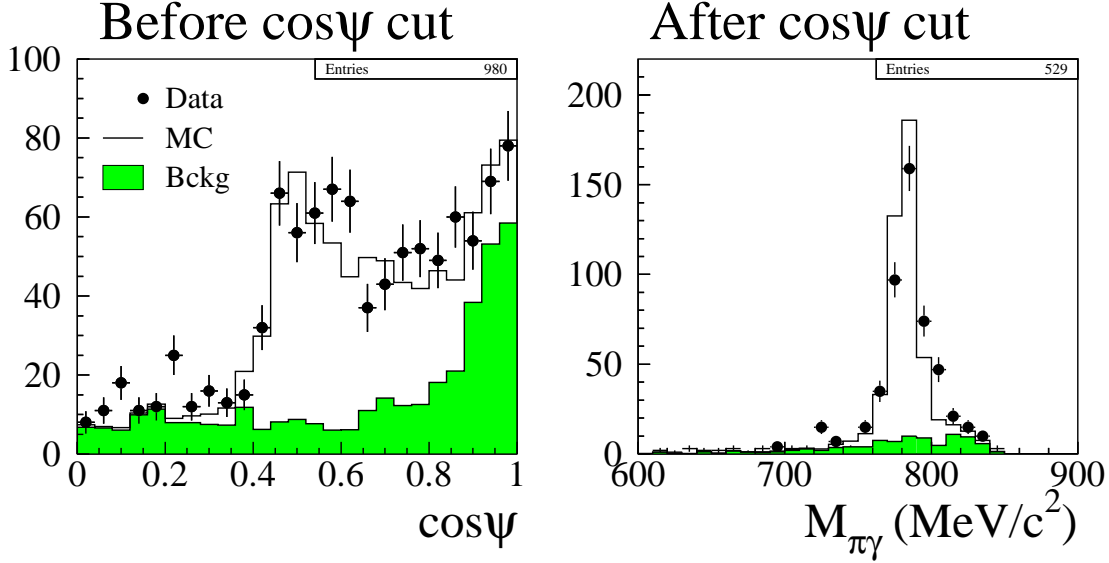,
   height=.35\textheight}
\end{center}
\caption{$\cos\psi$ variable before applying $\omega\pi^0_{\rm cut}$ (left)
         and $\pi^0\gamma$ invariant mass after this cut (right). 
         Fit constraint results have been used.
         In green the expected background contributions from Monte Carlo is
         superimposed while solid histograms are simulated distribution
         including signal summed up with the background.}
\label{Fig:Data_wp}
\end{figure}

The $f_0\gamma$ fit applied to data yields 679 events with a good $\chi^2$ out 
of which 307 survive also the $f_0\gamma_{\rm cut}$. 
In Figs.\ \ref{Fig:F0nEne}.left (center) the radiative photon's energy and the 
invariant mass of the $\pi^0\pi^0$ system are shown after kinematic fit for 
the events before (after) applying this cut. In the same distributions the 
expected background contribution, estimated from MonteCarlo, is reported in 
the solid coloured shapes; the S/B ratio after the cut improves of at least a 
factor 3, as expected, and a clear peak above background appears around 950 
MeV in the invariant mass (right). 

After subtracting the $112 \pm 11$ background events, the total counting
for the signal 
is $195 \pm 20$ (stat.). As in the $\omega\pi^0$ case, 8\% systematic error is 
assigned to the measurement. We obtain for $M_{\pi\pi}>700$ MeV:
\begin{equation}
{\rm BR}(\phi\to f_0\gamma\to\pi^0\pi^0\gamma) =
  (\,0.81\pm 0.09\,({\rm stat.})\pm 0.06\,({\rm sist.})\,)\times 10^{-4}
\end{equation}

\noindent
to be compared with the results of the Novosibirsk experiments\cite{cmd2_f0,snd_scal}.
%$Br(\phi\rightarrow\pi^0\pi^0\gamma)=(1.14\pm0.10\pm0.12)\cdot10^{-4}$ from
%SND and,
%$Br(\phi\rightarrow\pi^0\pi^0\gamma)=(1.06\pm0.09\pm0.06)\cdot10^{-4}$ from
%CMD-2.

This measurement is still to be considered preliminary since work is in 
progress to estimate the systematic errors directly from the data and to
correct the analysis efficiency as a function of $M_{\pi\pi}$ in order to
calculate a BR independently from any Monte Carlo assumptions.

\begin{figure}
\begin{center}
\begin{tabular}{cc}
 \epsfig{file=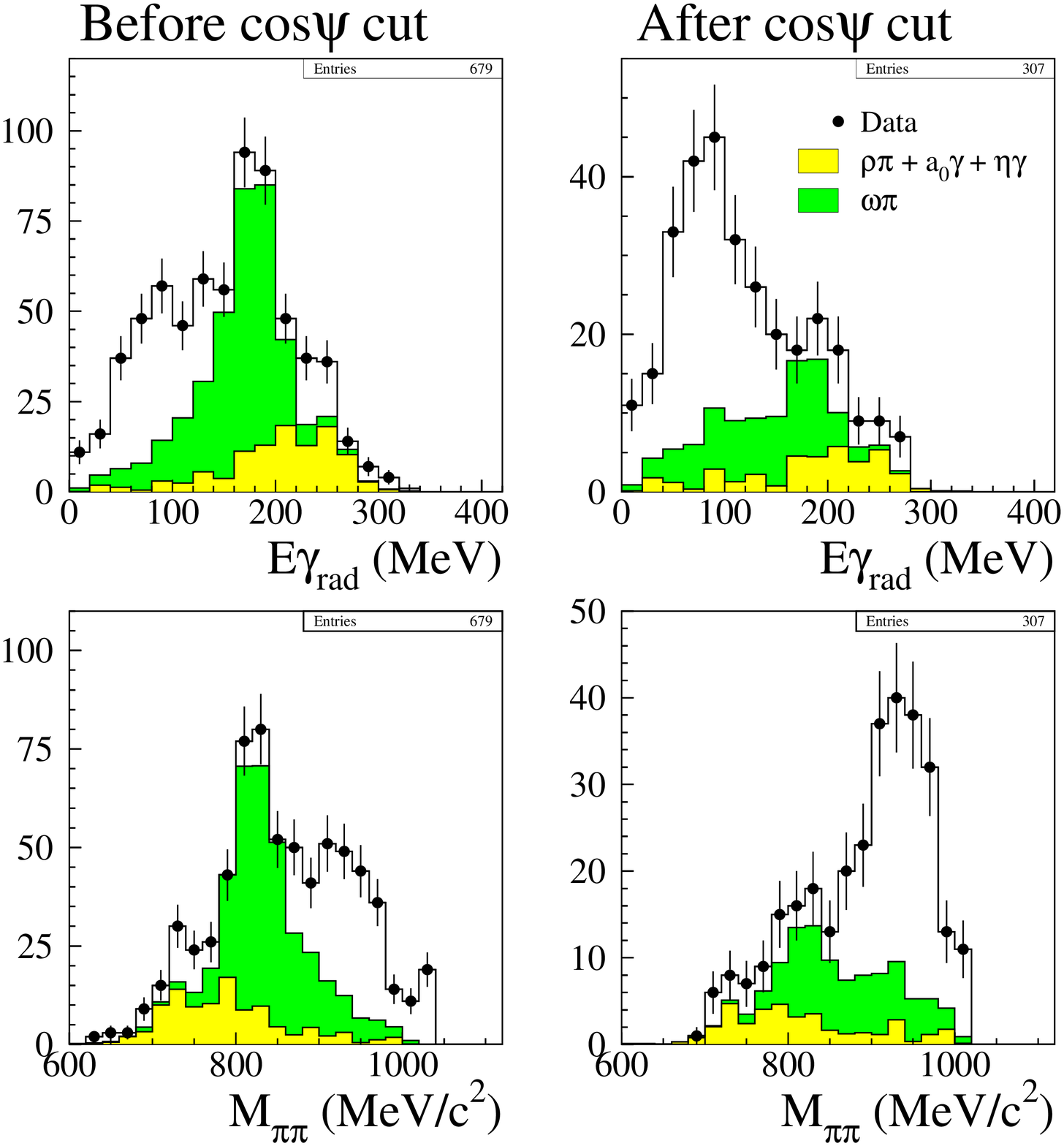,
       height=0.5\textheight}&
 \epsfig{file=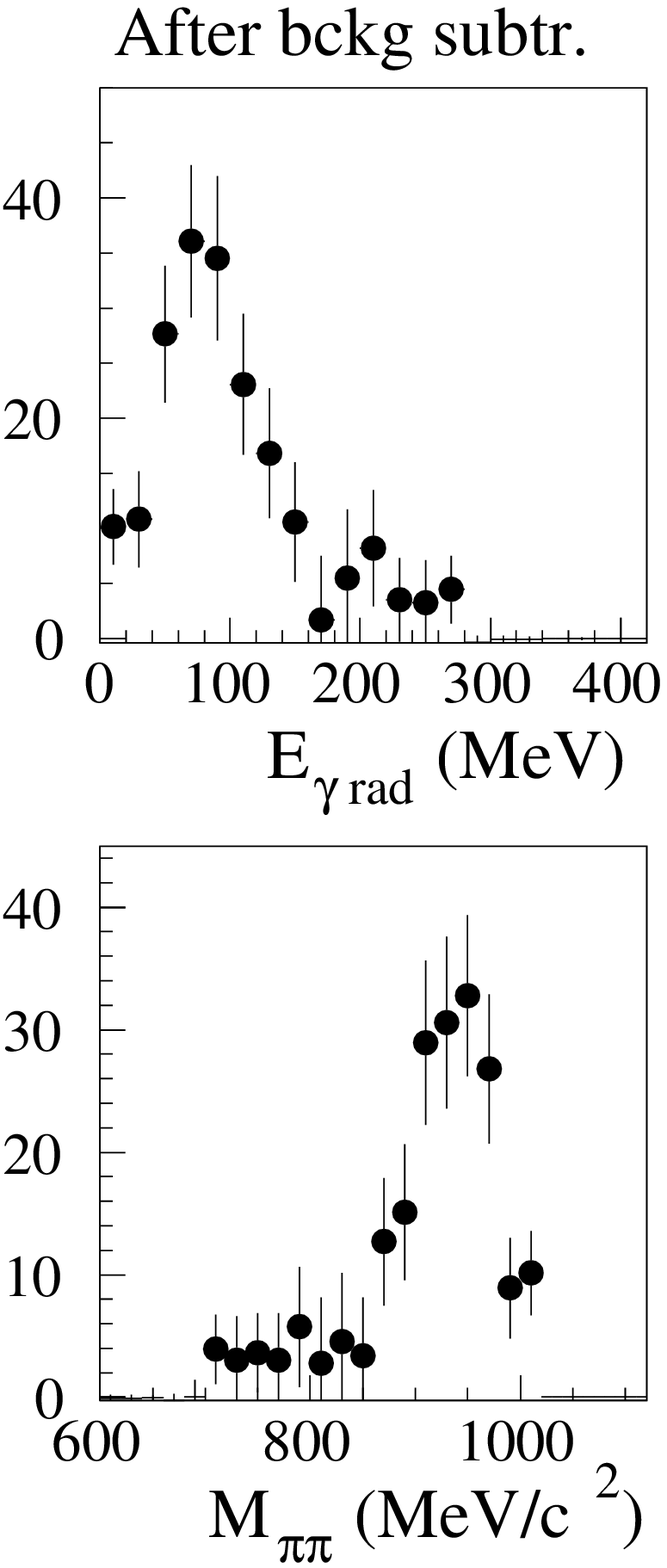,
       height=0.5\textheight}\\
\end{tabular}
\end{center}
\caption{Distributions of the radiative photon's energy and of the 
         $\pi^0\pi^0$ invariant mass before (left) and after (center) the
         $f_0\gamma_{\rm cut}$. The background contribution, estimated from 
         Monte Carlo, is superimposed. The green region represents 
         $\omega\pi^0$ events while in yellow all other contributions are 
         summed up. Right : same distributions after background 
         subtraction.}
\label{Fig:F0nEne}
\end{figure}

\subsection{$\phi\rightarrow f_0\gamma$, with
$f_0\rightarrow\pi^{+}\pi^{-}$}
\label{bvaleriani}

The analysis of the $\phi \rightarrow f_{0} \gamma$ decay in the charged
channel $f_{0} \rightarrow \pi^{+}\pi^{-}$ has been performed on a sample
of $1.8 \, {\rm pb}^{-1}$  
of collected data by looking at the spectrum of the production cross
section of $\pi\pi\gamma$ events 
as a function of $\pi^{+}\pi^{-}$ invariant 
mass squared, $Q^{2}$\cite{barbara}.
Two other processes contribute to the $\pi^{+} \pi^{-} \gamma$ final state:
{\em Initial State Radiation} 
(ISR), in which the photon is emitted by the incoming electron or positron,
and {\em Final State Radiation} (FSR), in which the $\gamma$ is emitted
by one of the  two pions. 
The latter process gives rise to an interference with the signal whose sign 
is not known. \\

The $\pi^{+} \pi^{-} \gamma$ events are selected using both drift chamber 
and electromagnetic calorimeter informations. The first step of the
signal selection requires a \emph{prompt} neutral cluster and
a vertex close to the interaction point. 
 This general selection identifies  
not only $\pi^{+} \pi^{-} \gamma$ events, but also 
$\mmg$ events and a huge amount of radiative Bhabhas.\\ 
The kinematical properties alone are not enough to suppress the $ee\gamma$
events, therefore a likelihood method has been developed
which uses both the particle time of flight and some informations coming 
from the cluster associated to the particle.
This method has a $95\%$ selection efficiency for a pion and 
a $\sim 94\%$ rejection power for electrons. \\

After the likelihood selection, 
kinematical cuts have been chosen in order to get a further reduction of
$\mmg$ and $\eeg$  
background and to emphasize the $\phi \rightarrow \pi^{+} \pi^{-} \gamma$ 
decay contribution. 
For both these purposes particles have been selected in the central part 
of the detector ($45^o<\theta<135^o$), since the 
polar angle distribution of the charged tracks from $\eeg$ and $\mmg$
events and of the photon  
from ISR are enhanced at small angles. \\

The last cut selects events based on the invariant mass of the charged
track identified by applying 4-momentum conservation in the hypothesis of
a massless neutral particle: 
$$
\left( \vec{p}_{1}+\vec{p}_{2} \right)^{2} - \left(
M_{\phi}-\sqrt{\vec{p}_{1}^{2}+M_{TR}^{2}} 
-\sqrt{\vec{p}_{2}^{2}+M_{TR}^{2}} \right)^{2} = 0
$$
where $\vec{p}_{1}$ and $\vec{p}_{2}$ are the tracks momenta and $M_{TR}$ 
is the track mass, assumed to be the same for both charged particles. 
The distribution of the variable $M_{TR}$ before and after the likelihood
cut is shown in fig.\ref{Mtrack}. The pion peak is clearly visible and
the signal events are selected in a window of $\pm$10 MeV around $M_\pi
= 140$ MeV, the central value of the fit.\\

\begin{figure}[p]
  \begin{center}
    \epsfig{file=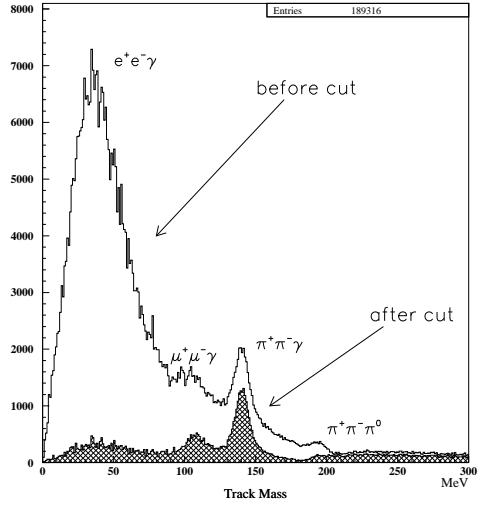,
                  height=.35\textheight}
\caption{Distribution of the variable $M_{TR}$ (see text) before and after
                  the likelihood selection. The pion and muon peaks are
                  clearly visible.}
\label{Mtrack}
  \end{center}
\end{figure}            

\begin{figure}[p]
  \begin{center}
    \epsfig{file=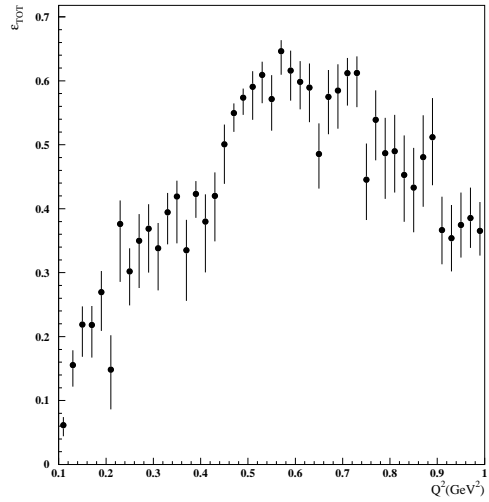,
                  height=.35\textheight}
\caption{Total selection efficiency for a $\pi^{+} \pi^{-} \gamma$ event in
the central part of the detector; the trigger  
         cosmic veto determines the low efficiency at high $Q^{2}$ values,
while the low $Q^{2}$  
         inefficiency is due to the kinematical cuts applied to reduce
$\ppp$ background.}  
\label{eff}
  \end{center}
\end{figure}      
\begin{figure}
  \begin{center}
    \epsfig{file=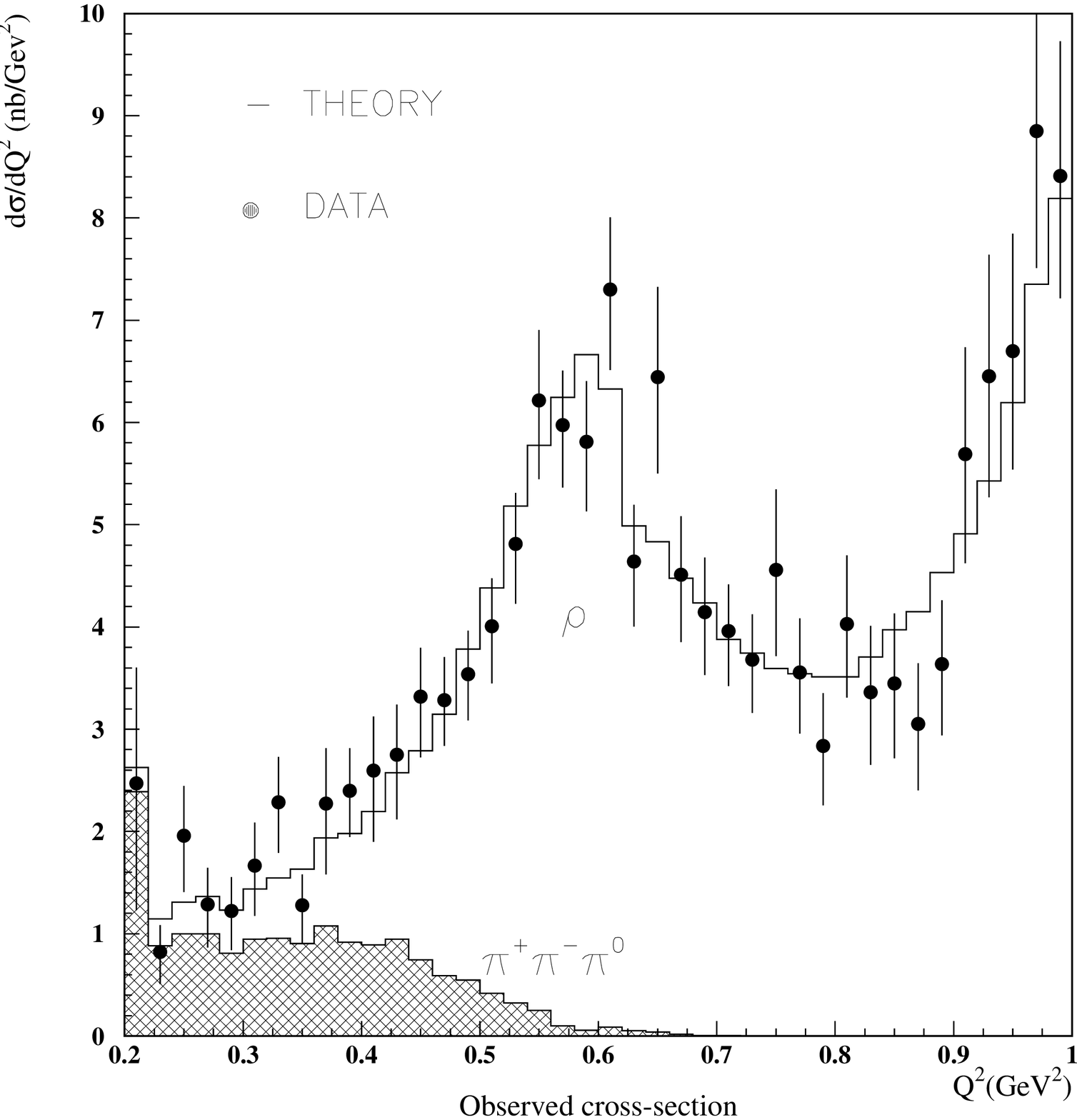,
                  height=.35\textheight}
\caption{Experimental cross-section as a function of $Q^{2}$ compared to
the theoretical one  
         for pure QED contributions.}
\label{cross-section_qq}
  \end{center}
\end{figure}

In order to compare the experimental results with the theoretical
predictions for $e^{+}e^{-}  
\rightarrow \pi^{+} \pi^{-} \gamma$ process, the $Q^{2}$ spectrum has been
corrected by the total selection  
efficiency as a function of $Q^{2}$, represented in fig. \ref{eff}. 
The low efficiency at high $Q^{2}$ values is due to the trigger {\em cosmic
veto}, which can  
mistakes a $\pi^{+} \pi^{-} \gamma$ event with a soft photon for a cosmic
event.\\ 
At low $Q^{2}$ values the efficiency decreases because of the cuts
applied to reduce $\ppp$ background. \\

Owing to the high rate of $\phi \rightarrow \ppp$ decays, $\ppp$ background
can survive $\pi^{+} \pi^{-} \gamma$  
selection: the contamination is higher at low $Q^{2}$ values, 
and it has been evaluated by fitting the track mass distribution 
of events identified as $\ppp$ for each $Q^{2}$ bin and extrapolating 
the fit function in the $\pi^{+} \pi^{-} \gamma$ mass region. \\

The experimental data, corrected by the total efficiency, have been fitted
using the theoretical  
spectrum including ISR and FSR contributions only.
The fit parameters are the integrated  
luminosity and the normalization factor of $\ppp$ spectrum with respect to
$\pi^{+} \pi^{-} \gamma$ one.
In fact to evaluate the $\pi^+\pi^-\pi^0$ background contribution to the
experimental differential cross-section, for $\pi^+\pi^-\pi^0$ events  has
been assumed the
same selection efficiency as for $\pi^+\pi^-\gamma$ ones, the latter fit
parameter (${\rm Norm}$) takes into account possible differences in the
overall efficiency. 

The results of the fit are:
 $L = 1774 \, {\rm nb}^{-1}$, ${\rm Norm} = 1.16$, with a
$\chi^{2}/ndf=33/45$.
Fig.\ref{cross-section_qq}  shows the comparison of the experimental
differential cross-section for the 
$e^{+}e^{-} \rightarrow \pi^{+} \pi^{-} \gamma$ process with the
theoretical one. The comparison is good and no $f_0$ signal is needed,
with the available statistics, to describe the spectrum. An upper limit
on the branching ratio for the decay $\phi \rightarrow f_{0}\gamma
\rightarrow \pi^{+} 
\pi^{-} \gamma$ can be set by fitting the $Q^{2}$ spectrum in the region
$Q^{2} < 0.84 \, {\rm GeV^{2}}$ 
and extrapolating it 
in the photon energy range $20 \, {\rm MeV} < E_{\gamma} < 120 \, {\rm
  MeV}$, where the signal is  expected. An excess of
 $35 \pm 160$ events, with
respect to the ones predicted by pure QED, is found. 
Assuming the isospin symmetry and ignoring the interference with FSR, 
this number corresponds to an upper limit on the 
value of the branching ratio of:
$$
            \begin{array}{ll} 
{\rm BR}(\phi \rightarrow f_{0}\gamma \rightarrow \pi^{+} \pi^{-} \gamma) <
1.64 \times 10^{-4} & \mbox{@ 90\% C.L.} \end{array}. 
$$

\subsection{$\phi\rightarrow\eta\pi^{0}\gamma$ with
$\eta\rightarrow\gamma\gamma$ } 

This process is characterized by 5 prompt photons without charged tracks in
the final state. 
It is expected to be dominated by the $\phi\rightarrow a_0\gamma$ decay, in
which the $a_0(980)$ decays into $\eta\pi^0$. 
The spectrum of the photon radiated by the $\phi$ is expected to be broad
and peaked at $\sim 50$ MeV.
Two other processes contribute to this final state:
$\phi\rightarrow\rho^0\pi^{0}$ and the non-resonant process
$e^+e^-\rightarrow\omega\pi^0$ with the rare decays of $\rho^0$ and
$\omega$ into $\eta\gamma$.
 
The main background comes from $\pi^{0}\pi^{0}\gamma$ final state, which
is dominant in the 5 photon sample; the expected number of events is 10
times bigger than the signal. 
The other relevant background comes from the $\phi\rightarrow\eta\gamma$
decay, with 3 and 7 photons in the final state, that can be reconstructed
as 5 photon events due to accidentals in the calorimeter or photon
splittings and mergings.
According to the MC the probability for both processes to be reconstructed
as 5 photon events is about 3\%, then due to their high branching ratio the
expected number of events in the 5 photon sample is also of the order of 10
times the signal.

The events are selected by requiring:
\begin{enumerate}
        \item no tracks in the drift chamber,
        \item total energy in the calorimeter greater than 900 MeV, 
        \item exactly 5 prompt photons in the angular acceptance region.
\end{enumerate}

On the 2200 events selected in the 2.4 pb$^{-1}$ sample, a first kinematic
fit has been applied by imposing constraints on the total energy and
momentum conservation, and on the consistency of time and position in the
calorimeter ($t-r/c=0$) for each photon. 
A cut corresponding to $P(\chi^2)<1\%$ has been applied.

Then for each event three different variables are constructed in order to
test the three hypotheses:

\begin{enumerate}
        \item $\eta\pi^0\gamma$ hypothesis:  
                $D_{\eta\pi^0\gamma}=\sqrt{\frac{(M_{12}-M_{\pi^0})^2}{\sigma^2_{\pi^0}}+
                              \frac{(M_{34}-M_{\eta})^2}{\sigma^2_{\eta}}}$
        \item $\pi^0\pi^0\gamma$ hypothesis:  
                $D_{\pi^0\pi^0\gamma}=\sqrt{\frac{(M_{12}-M_{\pi^0})^2}{\sigma^2_{\pi^0}}+
                              \frac{(M_{34}-M_{\pi^0})^2}{\sigma^2_{\pi^0}}}$
        \item $\eta\gamma$ hypothesis:  
                $D_{\eta\gamma}=\sqrt{\frac{(M_{12}-M_{\eta})^2}{\sigma^2_{\eta}}+
                              \frac{(E_3-E_{rad})^2}{\sigma^2_{rad}}}$
\end{enumerate}

The value of each $D$-variable is obtained by choosing the photon pairing
that minimizes it.
$M_{12}$ and $M_{34}$ are the invariant masses of the photon pairs,
$E_{rad}$=363 MeV.
$\sigma_{\eta}=20$ MeV and $\sigma_{\pi^0}=9$ MeV have been evaluated from
the data themselves, by fitting the invariant masses distribution on a
sample of events.
$\sigma_{rad}$ is obtained from the energy resolution of the calorimeter.

The following cuts are applied:

\begin{enumerate}
        \item $D_{\eta\pi^0\gamma} < D_{\pi^0\pi^0\gamma}$ in order to
              select the events that have a bigger probability to be
              $\eta\pi^0\gamma$ rather than $\pi^0\pi^0\gamma$ (see
              fig.\ref{a05g_d1}) 
        \item $D_{\eta\gamma} >$ 2; in fig.\ref{a05g_d2} is shown that this
              cut is able to reject events in the peak of $E_{rad}$
              at 363 MeV, that correspond to the $\eta$ mass peak.
\end{enumerate}

\begin{figure}
        \begin{center}
                \epsfig{file=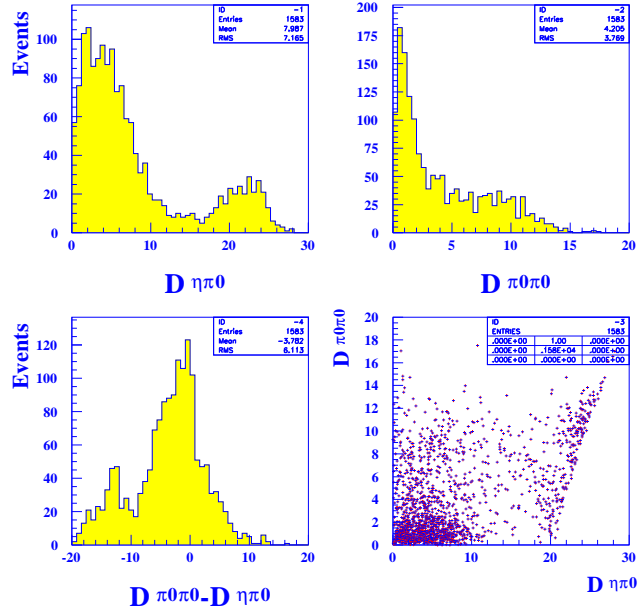,height=.35\textheight}
        \end{center}
        \caption{$D$-variables for the $\eta\pi^0\gamma$ and
                 $\pi^0\pi^0\gamma$ hypotheses; a cut is applied on their
                 difference by selecting the positive part of the
                 distribution.}  
        \label{a05g_d1}
\end{figure}    

\begin{figure}
        \begin{center}
                \epsfig{file=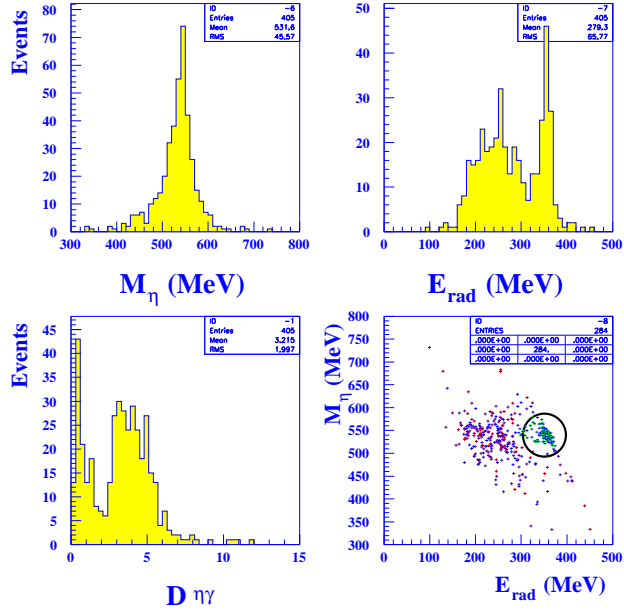,height=.35\textheight}
        \end{center}
        \caption{Rejection of $\eta\gamma$ background, the circle in the
                 scatter plot shows the rejected region with the cut
                 $D_{\eta\gamma}>2$.} 
        \label{a05g_d2}
\end{figure}    

On the 240 events selected a second kinematic fit is applied, imposing as
further constraints the two invariant masses of $\eta$ and $\pi^0$; 
a cut corresponding to $P(\chi^2)<1\%$ has been applied, 
and the 74 surviving events form the final sample. 

The efficiencies for signal and backgrounds have been
evaluated by MC, taking into account the dependence on photon energy.
The resulting efficiencies are listed in Tab.\ref{a05g_eff}:
in the first column is reported the trigger plus the background filter
one; in the second column, the efficiency of the 5 prompt photon
cut, that is mostly due to the angular cut at $21^o$, and in third one is
reported the effect of the selection and of the two kinematic fits.

\begin{table}
        \begin{center}
                \begin{tabular}{|c|c|c|c|c|}
\hline
        Process & Trigger + & 5 prompt & Selection + & Total \\
                & filters   & photons  & kin. fits   & \\
\hline 
        $\phi\rightarrow a_0\gamma\rightarrow\eta\pi^0\gamma$      & 0.81 &
0.70 & 0.40 & 0.23 \\
        $\phi\rightarrow\rho^0\pi^{0}\rightarrow\eta\pi^0\gamma$   & 0.77 &
0.70 & 0.26 & 0.14 \\
\hline 
        $\phi\rightarrow f_0\gamma\rightarrow\pi^0\pi^0\gamma$     & 0.80 &
0.70 & 2$\cdot 10^{-3}$ & $10^{-3}$ \\
        $\phi\rightarrow\rho^0\pi^{0}\rightarrow\pi^0\pi^0\gamma$  & 0.80 &
0.70 & 0.03 & 0.02 \\
        $e^+e^-\rightarrow\omega\pi^0\rightarrow\pi^0\pi^0\gamma$ & 0.82 &
0.70 & 0.02 & 0.01 \\
\hline
        $\phi\rightarrow\eta\gamma$             & 0.78 &
0.03 & -    & $< 5\cdot 10^{-4}$ \\
\hline
                \end{tabular}
        \end{center}
        \caption{Efficiencies evaluated by means of the MC simulation; for
                 the
                 $e^+e^-\rightarrow\omega\pi^0\rightarrow\eta\pi^0\gamma$
                 process we assume the same efficiencies of the
                 $\pi^0\pi^0\gamma$ final state.}  
        \label{a05g_eff}
\end{table}             

\begin{figure}[p]
        \begin{center}
                \epsfig{file=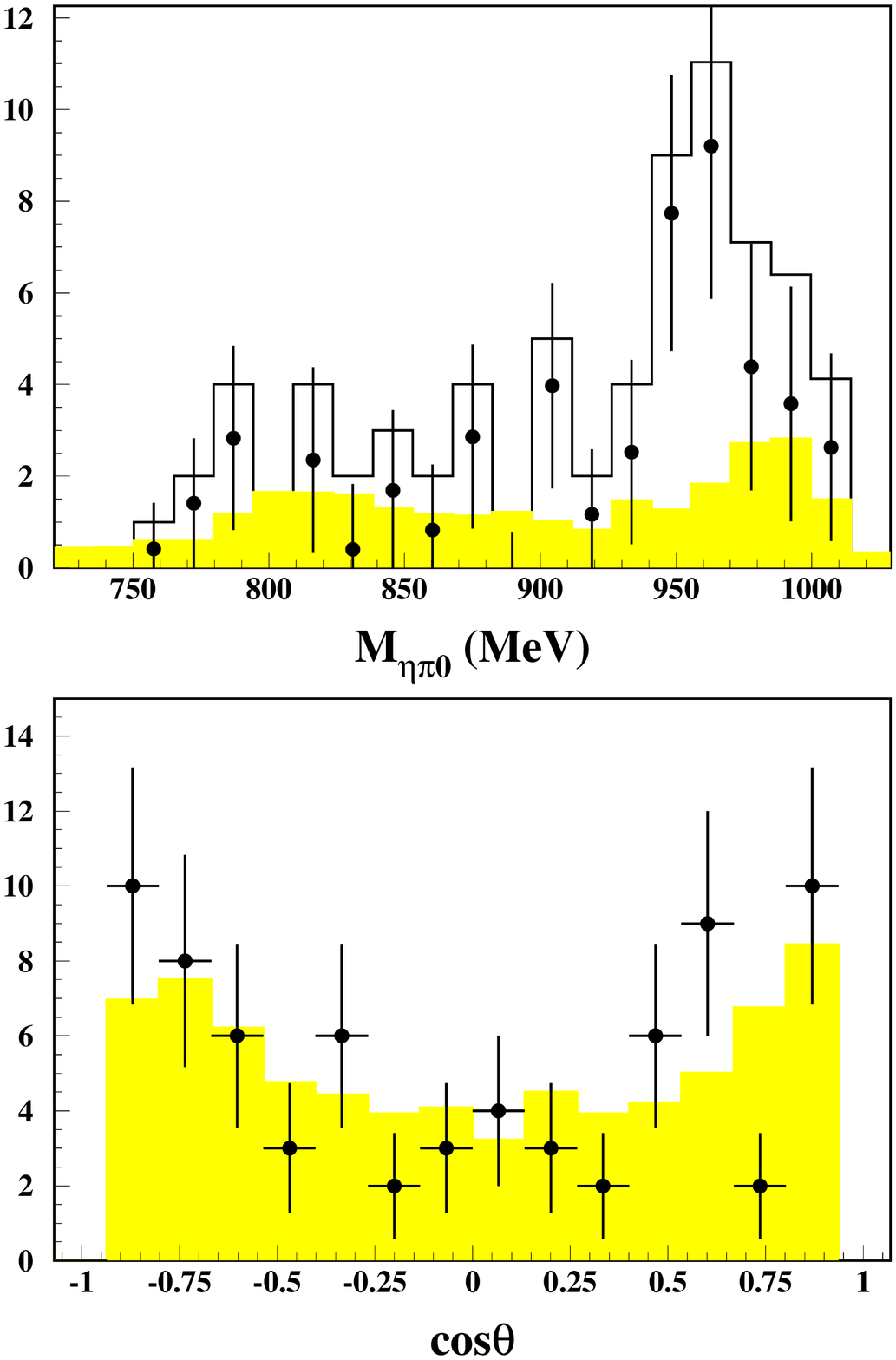,height=.5\textheight} 
        \end{center}
        \caption{Upper plot: invariant mass of the $\eta\pi^0$ system,
                 histogram: data; shadowed histogram: expected background 
                 (from MC); black points: data - background difference.
                 Lower plot: distribution of the cosine of the polar angle
                 of the radiated photon, black points: data; histogram: MC
                 - signal only.}    
        \label{a05g_spectrum}
\end{figure}    

In fig.\ref{a05g_spectrum} are reported the invariant mass spectrum of the
four photons assigned to $\eta$ and $\pi0$ and the distribution of the
cosine of the polar angle of the radiated photon, which agrees with the
expected $1+cos^2\theta$.

The background spectrum, superimposed in fig.\ref{a05g_spectrum}, is a 
prediction obtained by MC by weighting the background processes
with the product of their expected branching ratio times their efficiency.   

In order to evaluate the $Br(\phi\rightarrow\eta\pi^0\gamma)$ we consider the
whole spectrum: 74$\pm$ 9 events are selected ($N$), with an expected
background of 21$\pm$ 6 ($B$), and assuming
$Br(\eta\rightarrow\gamma\gamma)=39.2\%$ 
\begin{equation}
        Br(\phi\rightarrow\eta\pi^0\gamma)=
        \frac{N-B}{\varepsilon L\sigma_{\phi}Br(\eta\rightarrow\gamma\gamma)}=
        (0.77\pm 0.15_{stat}\pm 0.11_{syst})\cdot 10^{-4}
\end{equation}
where $L=2.4$ pb$^{-1}$, and $\sigma_{\phi}=3.2$ $\mu$b.

This value is in agreement within the errors with the results of the
Novosibirsk experiments\cite{snd_scal,cmd2_f0}.

According to the MC we expect 6 events from the processes
$\phi\rightarrow\rho\pi^0\rightarrow\eta\pi^0\gamma$ and
$e^+e^-\rightarrow\omega\pi^0\rightarrow\eta\pi^0\gamma$, then considering
them as background we obtain:
\begin{equation}
        Br(\phi\rightarrow a_0\gamma\rightarrow\eta\pi^0\gamma)=
        (0.69\pm 0.14_{stat}\pm 0.10_{syst})\cdot 10^{-4}
\end{equation}

\subsection{$\phi\rightarrow\eta\gamma$, with $\eta\rightarrow\gamma\gamma$}

The $\phi\to\eta\gamma\to\gamma\gamma\gamma$ decay, having an higher BR 
(0.49\%) and an harder energy spectrum with respect to the rest of $\phi$ radiative 
decays, is a good calibration sample for multi--photon final states. 

Since the energy of the radiative photon ($E_{\gamma_{\rm rad}} \sim 360$ 
MeV) is inside the energy range of the $\gamma$'s coming from the $\eta$ 
($150<E_{\gamma_\eta}<500$ MeV), the kinematic fit is used to select the 
$\gamma\gamma$ pair assigned to the meson.
For each event with three photons in time window the fit is applied in the 
$\eta\gamma$ hypothesis  using the $\eta$ mass constraint for the three 
possible photons' combination. The minimum $\chi^2$ is then selected.

The only relevant background for this channel comes from the 
$\phi\to\pi^0\gamma$ decay (${\rm S/B}\sim 4$). Since the energy of the 
primary photon ($E_{\gamma_{\rm rad}} \sim 500$ MeV) is higher than the one
of the signal, the fit can reconstruct it as an $\eta\gamma$ event 
by wrongly combining the radiative photon with a $\gamma$ coming from the 
$\pi^0$ ($E_{\gamma_{\pi^0}}<500$ MeV). Being the third photon constrained 
in the 360 MeV region, the three photons are monochromatic for $\pi^0\gamma$
events after the fit.
The background is therefore rejected using the $\Delta E$ variable, energy 
difference of the $\gamma\gamma$ pair, requiring $|\Delta E|<330$ MeV:
as it is shown in Fig.\ \ref{Fig:DataSummary_etag}.up-left, 
$\pi^0\gamma$ events are peaked at high $|\Delta E|$ values while signal has 
a flat distribution.

A sample corresponding to 1.84 pb$^{-1}$ have been analysed: among 226736 
events with at least 3 clusters, 183345 satisfy angular acceptance cut and 
45889 have also 3 photons in time window with $E>20$ MeV. 
After the $\chi^2$ cut and the background rejection the sample is reduced to
18504 events.
The resulting angular and energy distributions of the radiative photon, 
together with the $\gamma\gamma$ pair invariant mass, are shown in Fig.\
\ref{Fig:DataSummary_etag}. 
All distributions are well in agreement with Monte Carlo expectations.
The resulting $\eta$ mass is within 0.2\% the expected value.
\begin{figure}
\begin{center}
\epsfig{file=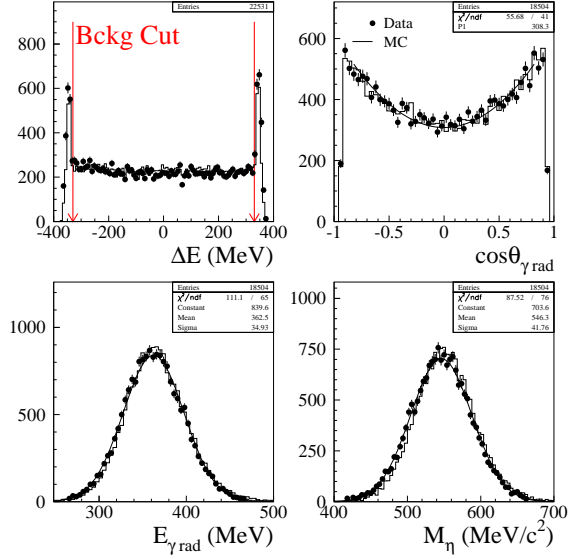,
  height=.35\textheight}
\end{center}
\caption{Comparison between data (---) and Monte Carlo ($\bullet$)
distributions 
  for $\phi\to\eta\gamma\to\gamma\gamma\gamma$ events: energy difference of
the 
  two photons assigned to $\eta$ (up-left) -- the two peaks are due to the
  $\phi\to\pi^0\gamma$ background; angular distribution (up-right) and energy 
  spectrum (down-left) of the radiative photon; $\gamma\gamma$ pair invariant 
  mass (down-right). Last two variables are obtained after photons' assignment 
  from the fit but using energies reconstructed without constraints.}
\label{Fig:DataSummary_etag}
\end{figure}

In order to check if $\gamma\gamma\gamma$ QED background can simulate the 
signal, angular distributions between photons' pairs have been studied 
(Fig.\ \ref{Fig:Angles_etag}). The excellent agreement with $\eta\gamma$ 
simulated events leaves no room for any residual background.
\begin{figure}
\begin{center}
\epsfig{file=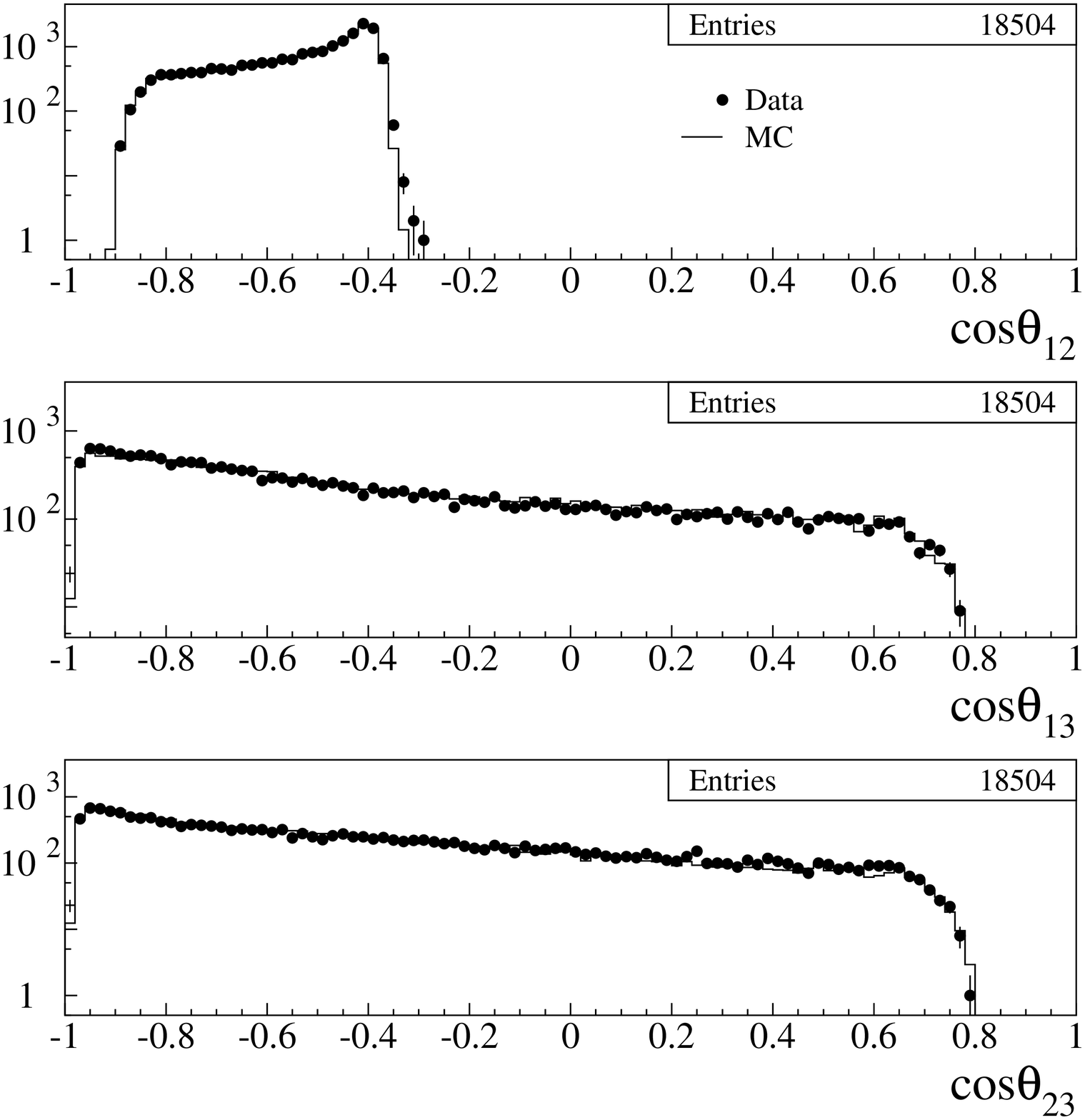,
  height=.35\textheight}
\end{center}
\caption{Opening angles between photons' pairs of 
  $\phi\to\eta\gamma\to\gamma\gamma\gamma$ events for data and Monte Carlo. 
  Photons named 1 and 2 are the ones assigned to $\eta$.}
\label{Fig:Angles_etag}
\end{figure}

The $\phi$ visible cross section has been evaluated by measuring
\begin{equation}
\sigma(e^+e^-\to\phi\to\eta\gamma\to\gamma\gamma\gamma) = 
   \frac {N_{\eta\gamma}} { L \times \varepsilon_{\rm ana}} \cdot 
   \frac{1}{\rm BR(\phi\to\eta\gamma\to\gamma\gamma\gamma)}
\end{equation}
and using ${\rm BR}(\phi\to\eta\gamma\to\gamma\gamma\gamma)=(0.49\pm 0.02)\%$.
Luminosity is estimated by using large angle Bhabha's ($\theta>45^{\circ}$) while 
analysis efficiencies, listed in Tab.\ \ref{Tab:Eff_etag},
\footnote{The values of the
contributions to  $\varepsilon_{\rm cl}$ are the following:
 92.6\% from trigger, 81.5\% from background rejection, 97.0\% from event 
classification.}
are evaluated from Monte Carlo.
\begin{table}
\caption{Analysis efficiencies for $\phi\to\eta\gamma\to\gamma\gamma\gamma$ 
events.}
\begin{center}
\begin{tabular}{|c|c|} \hline
Efficiency                       &  Contribution                    \\ \hline
$\varepsilon_{\rm cl}=76.0\%$    &  
  Trigger, background rejection and event classification            \\
$\varepsilon_{\rm sel}=93.7\%$   &  Acceptance and time window cuts \\
$\varepsilon_{\chi^2}=98.5\%$    &  $\chi^2$ cut                    \\
$\varepsilon_{\Delta E}=91.3\%$  &  $\Delta E$ cut                  \\ \hline
$\varepsilon_{\rm ana}=64.3\%$   &  Total                           \\ \hline
\end{tabular}
\end{center}
\label{Tab:Eff_etag}
\end{table}

We quote:
\begin{equation}
\sigma_{\phi} = 
   (\,3.19\pm 0.02\,({\rm stat.})\pm 0.23\,({\rm syst.})\,)\ \mu{\rm b}
\end{equation}
This value is in good agreement with CMD-2 result, obtained using 
$K_S\to\pi^+\pi^-$ events: 
$\sigma_{\phi} = (3.114\pm 0.034\pm 0.048)\ \mu{\rm b}$\cite{CS_CMD2}.

Complete evaluation of systematics is in progress. At the moment a preliminary 
estimate of the contributions is summarized in Tab.\ \ref{Tab:Syst_etag}. 

\begin{table}
\caption{Error contributions to $\phi$ cross section measurement using 
$\phi\to\eta\gamma\to\gamma\gamma\gamma$ events.}
\begin{center}
\begin{tabular}{|c|c|c|} \hline
VARIABLE                    &  VALUE    &  METHOD        \\ \hline
BR$_{\gamma\gamma\gamma}$   &  $4\%$    &  From PDG      \\ \hline
$L$                         &  $3\%$    &  Comparing trigger vs L3 vs offline 
                                           values        \\ \hline
$\varepsilon_{\rm cl}$      &  $5\%$    &  
  \begin{tabular}{c}
    Conservative estimate using a small data \\
    control sample without event classification \\ 
  \end{tabular} \\ \hline
$\varepsilon_{\rm sel}$     &  $1.0\%$  & 
  \begin{tabular}{c}
    Fit to $\cos\theta_{\rm\gamma\,rad}$ distribution in different angular\\
    regions for acceptance. No contributions from\\
    TW cut (TW$_{\rm max}\ 2\to 3$ ns: no changes) and\\
    clustering ($\varepsilon_{\rm MC}=99.9\%$)\\
  \end{tabular} \\ \hline
$\varepsilon_{\chi^2}$      &  $0.6\%$  &  
  \begin{tabular}{c}
    $50\%$ of the wrong assignments percentage as \\
    obtained from MC\\
  \end{tabular} \\ \hline
$\varepsilon_{\Delta E}$    &  $0.3\%$  &   
  \begin{tabular}{c}
    Comparison between loss of events counted\\
    with MC and the estimate on data fitting the\\
    $|\Delta E|< 300$ MeV region\\
  \end{tabular} \\ \hline
\end{tabular}
\end{center}
\label{Tab:Syst_etag}
\end{table}

In addition to the analysis described above an alternative method has been
developed for measuring the ratio
Br($\phi\rightarrow\eta\gamma$)/Br($\phi\rightarrow\pi^0\gamma$) without 
using any kinematic fit\cite{scuri}.

\subsection{The ratio BR(\fietapg)/BR(\fietag)}
%\subsection{The pseudoscalar sector: \fietapg,  \fietag}

Two different decay chains giving rise to both fully neutral and
charged/neutral final states
have been used to study the ratio $R_{\phi}= BR(\fietapg)/BR(\fietag)$:
\begin{enumerate}
\item $\fietapg \rightarrow \eta\pip\pim\gamma\rightarrow
\pip\pim\gamma\gamma\gamma $
\item $\fietapg \rightarrow \eta\piz\piz\gamma \rightarrow 7\gamma$
\end{enumerate}
A very clean control sample is given by \fietag
decays with identical final state:
\begin{enumerate}
\item $\fietag \rightarrow \pip\pim\piz\gamma\rightarrow
\pip\pim\gamma\gamma\gamma$
\item $\fietag \rightarrow \piz\piz\piz\gamma\rightarrow 7 \gamma$
\end{enumerate}
These events, being 2-3 order of magnitude more probable than the
corresponding \fietapg ones, constitute also the main source of background
for their detection. For this reason a kinematic fit with mass constraints
is needed to obtain a satisfactory signal to background ratio in these
final states. 

Since the final state is identical for
the \fietapg and \fietag corresponding channel, most of the
systematics will cancel if we evaluate the ratio $R_{\phi}$ using the same
final state to count \fietag and \fietapg events.

\subsubsection{\pip\pim\phot\phot\phot final state}
For \fietapg events this final state is characterized by a nearly
monochromatic photon with $E_{\gamma}=60 \MeV$ recoiling against the \etap,
and two (harder) photons coming from \etagg annihilation.
On the contrary, for \fietag events the radiative photon, still
monochromatic, is the most energetic 
one, with $E_{\gamma}=363 \MeV$ and the two (softer) other photons come
from \piz$\rightarrow\gamma\gamma$ annihilation.
  
In addition to the \fietag background, some background events can be
expected from \fikskl events with one charged and one neutral vertex where
at least one photon is lost and the \Kl is decaying near the interaction
point and from $\phi\rightarrow \pip \pim \piz$ events with an additional
cluster counted.  

A first level {\em topological} selection runs as follows:
\begin{itemize}
\item 3 and only 3 prompt neutral clusters (as described above) with
$E_{\gamma}>10 \MeV$ and $21^{\circ}<\theta_{\gamma}<159^{\circ}$
\item 1 charged vertex inside the cylindrical region $r<4\cm$; $|z|<8 \cm$
\end{itemize}
and is common for both \fietapg and \fietag events.

Background from \pip\pim\piz events is strongly reduced by means of 
a cut on the sum of the energies of the charged tracks assumed to be pions:
it is expected larger for three pions events than for radiative events.

Background from \fikskl is reduced for \fietag using the fact that the
spectrum for photons coming from kaons is limited to
energies below 280 MeV while in \fietag  we expect at least one photon with
energy exceeding 300 MeV. 

The same cut cannot be applied to \fietapg events
where the energy spectrum of photons is different: in this case
however, a suitable variable to select the signal has been found to be the
sum of the energies of the three photons $E_{\gamma\gamma\gamma}$.
In conclusion one applies the cuts:

\begin{itemize} 
\item For \fietag selection:
        \begin{itemize}
        \item[-] $E_{\pip}+E_{\pim} < 550 \MeV$ 
        ($\varepsilon_{3\pi}\simeq 1.5\cdot 10^{-3}$)
        \item[-] $E_{\gamma}^{\rm max} > 300 \MeV $     
        ($\varepsilon_{\Kl\Ks}\simeq 2\cdot 10^{-4}$)
\end{itemize}
\item For \fietapg selection:
        \begin{itemize}
        \item[-] $E_{\pip}+E_{\pim} < 412 \MeV$ 
        ($\varepsilon_{3\pi}\simeq 1 \cdot 10^{-4}$)
        \item[-] $E_{\gamma\gamma\gamma} > 520 \MeV$ 
        ($\varepsilon_{\Kl\Ks}\simeq 1 \cdot 10^{-4}$)
        \end{itemize}
\end{itemize}
The efficiency for this selection is
evaluated from Monte Carlo to be 39.6\% for \fietapg and 37.9\% for \fietag
events.

Contamination from \fietag events into the \fietapg sample is at this level
still very high (S/B $\sim 10^{-3}$): thus a kinematic fit with
mass constraints (see section \ref{kinefit}) has been implemented for both
the decay chain hypotheses constraining all intermediate masses.
The energy spectrum of the photons gives no problem in assigning clusters to
particle originating them:
as already noticed above radiative photon is the most energetic one in
\fietag events, while it is 
the less energetic one in \fietapg events; the other two cluster belong to
\piz and \Eta respectively in the two cases.
A cross cut on ${ \cal P}(\chi^2_{\etap\gamma}) > 25\%$ {\em and}   
 ${ \cal P}(\chi^2_{\eta\gamma}) < 1\%$ has proven by Monte Carlo to
maximize the 
significance $S/\sqrt{B}$ and has thus been chosen as final selection
criterium for \fietapg events. Final Monte Carlo efficiency after this cut
is 18.6\% for \fietapg while for \fietag a 90\% C.L. upper limit can be set
to $4.4 \cdot 10^{-5}$ giving rise to an expected S/B ratio $> 35$ (90\% C.L.)
if one uses the PDG`98 value for BR(\fietapg).

A selection cut can also be put to select \fietag events, and has been
chosen in a very conservative way to be ${ \cal P}(\chi^2_{\eta\gamma}) >
1\%$ due to the low background on this channel.
With this cut one has a
final efficiency of 31.9\% for \fietag and selects a very pure set of
events with background (estimated from Monte Carlo, and confirmed by a fit
to the \Eta mass peak) being $\sim 0.1\%$ of 
the sample.

The abundant and pure  \fietag events can be used as control sample for
systematic effects on the efficiency, and to compare data versus Monte Carlo
distributions for the variable on which the cuts are set.
%%%%%%%%%%%%%%%%%%%%%%%%%%%%%%%%%%%%%%%%%%%%%%%%%%%%%
\begin{figure}
\begin{center}
\begin{tabular}{cc}
\epsfig{file=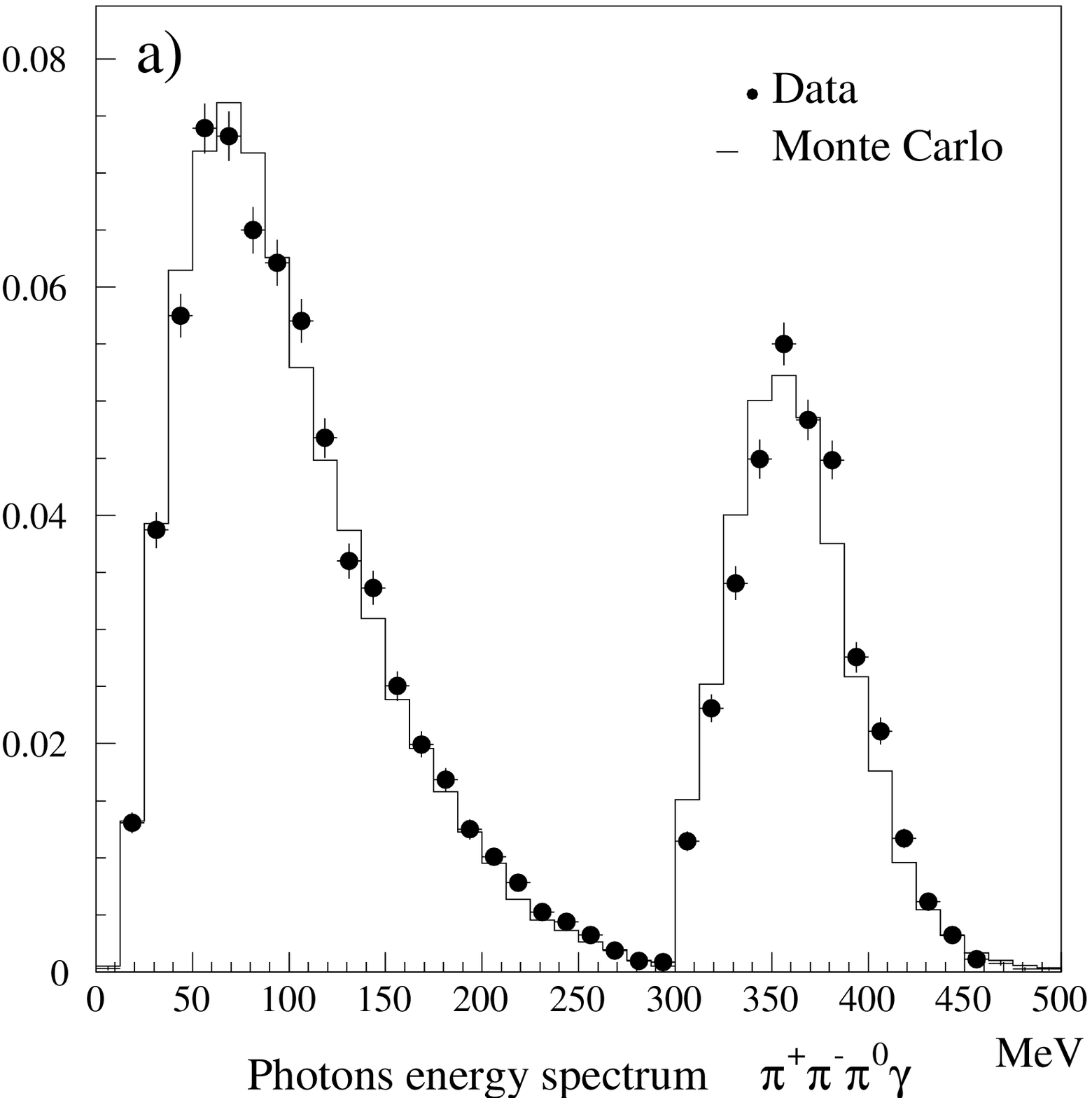,width=48mm}
&
\epsfig{file=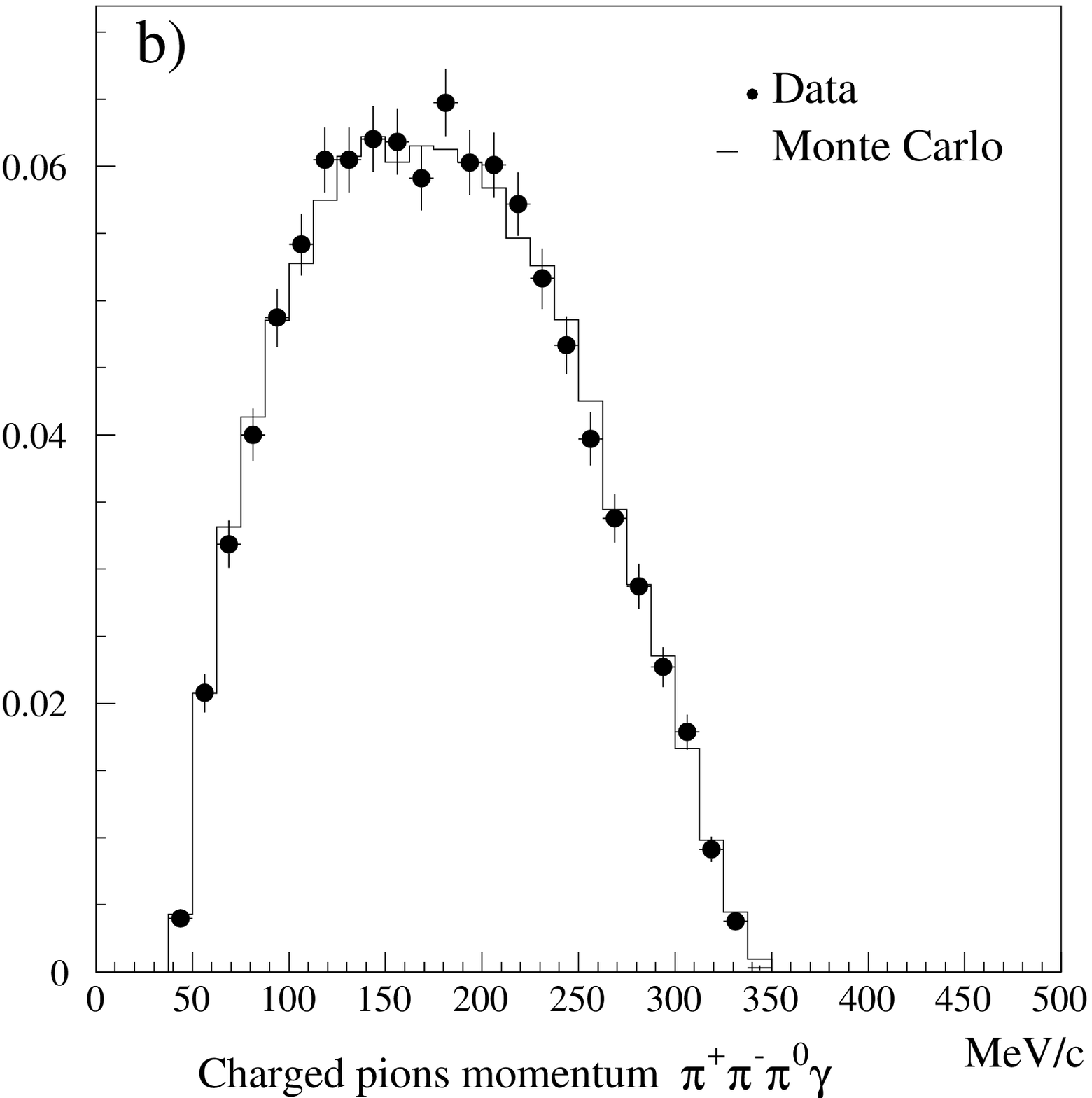,width=48mm}\\
\epsfig{file=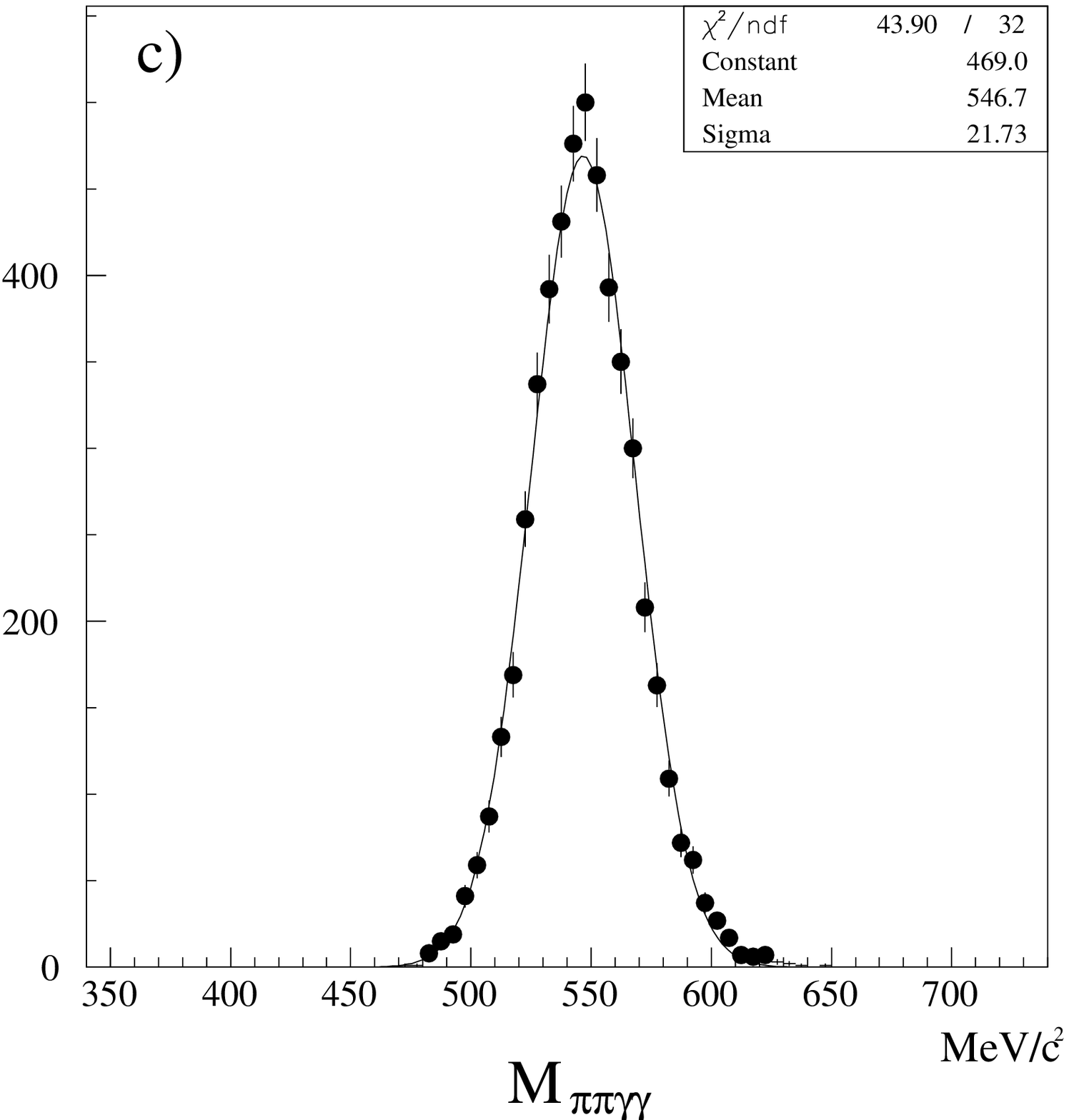,width=48mm}
&
\epsfig{file=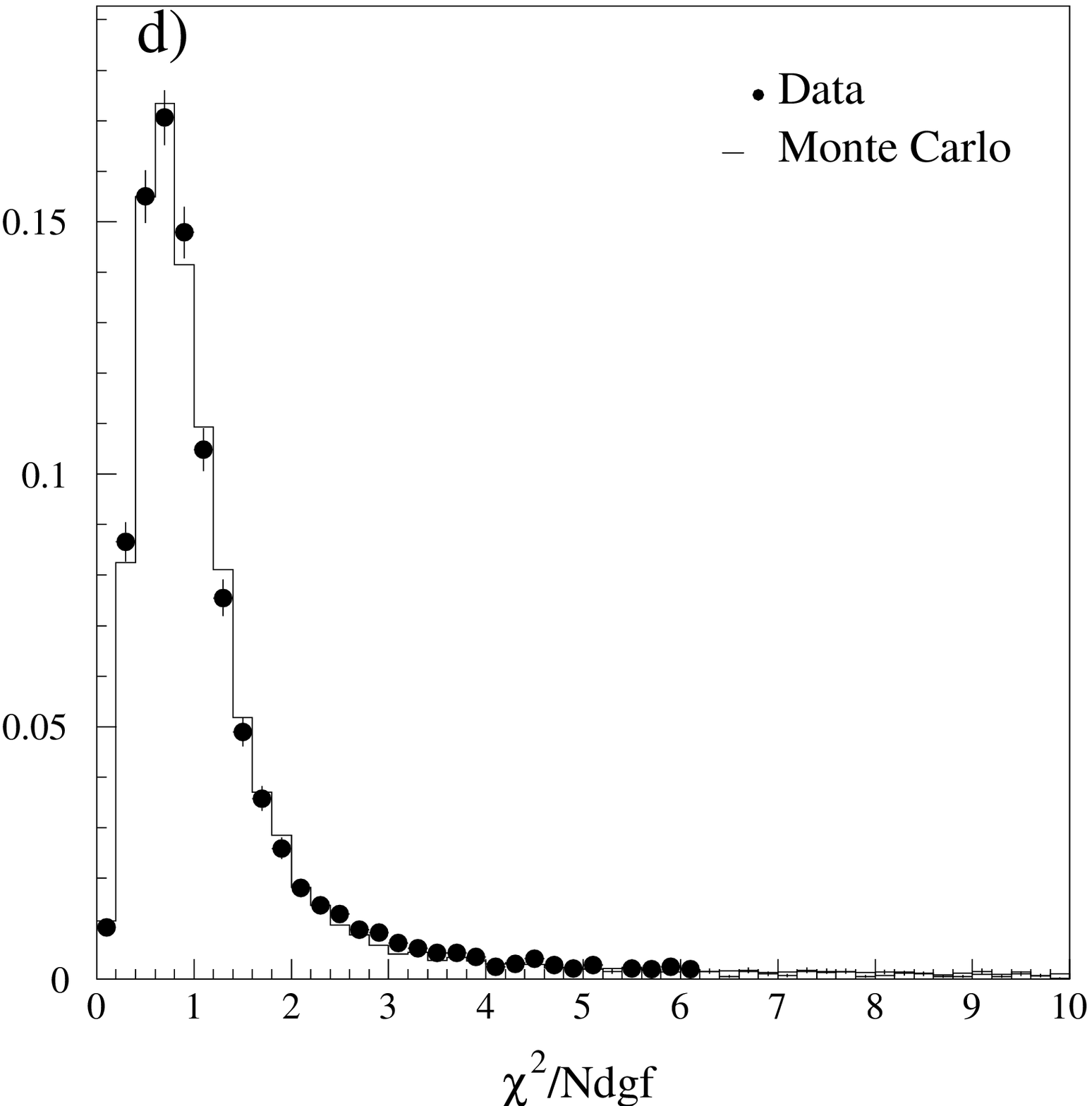,width=48mm}\\
\end{tabular}
\end{center}
\caption{{ Monte Carlo (pure \fietag) versus data for
    \pip\pim\phot\phot\phot events:
a) Cluster energy spectrum; b) Charged pions momentum spectrum; c) \Eta
invariant mass distribution; d) $\chi^2$ of kinematic fit}}
\label{fig_datavsmc_chneu}

\end{figure}
%%%%%%%%%%%%%%%%%%%%%%%%%%%%%%%%%%%%%%%%%%%%%%%%%%%%%

All comparisons (see fig.\ref{fig_datavsmc_chneu}) show very good agreement
(within 1-2\%) between 
data and Monte Carlo, and since the dependence of the efficiency on the
cuts is not critical (for example moving the cut on the charged pions
energy by $\pm 1\%$ changes \fietag selection efficiency by $\sim 0.1\%$,
for a more detailed discussion see \cite{tesiPhDFabio})
the overall systematic error on the estimation of efficiencies
is very small. Also, when evaluating the ratio $R_{\phi}$ most of the
systematics will cancel out due to the strong similarities between the two
categories of events.
With the statistics of $\sim 2.4 \pbinv$ of 1999 run we found 21$\pm$ 4.6
\fietapg events in this decay chain with less than one event of background
expected at 90\% 
C.L., while with the \fietag selection selects 6696 events in the same runs.
The distribution of the invariant mass of the two charged pions and the two
most energetic photons in the event is shown in fig. \ref{etapinvmass}
compared to the Monte Carlo expected for pure \fietapg events.
%%%%%%%%%%%%%%%%%%%%%%%%%%%%%%%%%%%%%%%%%%%%%%%%%%%%%
\begin{figure}
\begin{center}
\epsfig{file=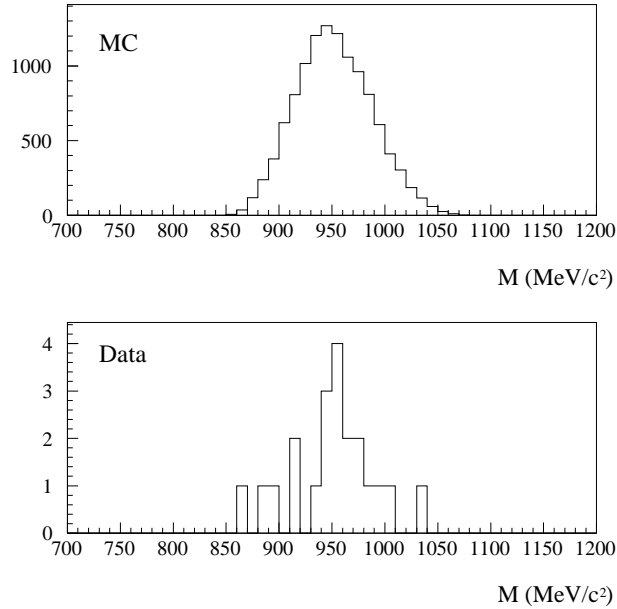,height=.35\textheight}
\end{center}
\caption{{ Invariant mass of two charged tracks and 
    two most energetic photons for events selected as \fietapg. Data (lower
    plot) is compared to pure monte Carlo \fietapg events.}}
\label{etapinvmass}

\end{figure}
%%%%%%%%%%%%%%%%%%%%%%%%%%%%%%%%%%%%%%%%%%%%%%%%%%%%%
Solving for $R_{\phi}$ we get:
$$
R_{\phi} =
\frac{N_{\etap\gamma}}{N_{\eta\gamma}}\frac{\varepsilon_{\eta\gamma}}{\varepsilon_{\etap\gamma}}\frac{BR(\etapippimpiz)BR(\piz\rightarrow\gamma\gamma)}{BR(\etappippimeta)BR(\etagg)}
$$
and, thus:
$$
R_{\phi} = \left(7.1 \pm 1.6 (\rm stat.) \pm 0.3 (\rm syst.)\right)\cdot
10^{-3}  
$$
where the systematic error is dominated
by the uncertainty on the value of the intermediate branching ratios (4\%).
In fact systematic effects on luminosity evaluation and  $\sigma_{\phi}$ cancel
out exactly in the ratio, while trigger efficiency, streaming efficiency
and reconstruction effects cancel almost exactly.
This result on $R_{\phi}$ leads in turn to: 
$$
BR(\fietapg) = \left(8.9\pm 2 (\rm stat.) \pm 0.6(\rm syst.)\right)\cdot 10^{-5}
$$ 
This result has to be compared to the most recent results by CMD-2 and SND,
which use the same final state\cite{etapsnd,etapcmd2}
%$(8.2^{+2.1}_{-1.9} {\rm stat.} \pm 1.1)\cdot 10^{-5}$ and 
%$(6.7^{+3.4}_{-2.9} {\rm stat.} \pm 1.0) \cdot 10^{-5}$ .

\subsubsection{7\phot final state}
The high hermeticity of the KLOE EmC allows us to detect with high efficiency
multiphoton final states. The
\fietapg$\rightarrow\piz\piz\eta\gamma\rightarrow 7\gamma$ and its
background \fietag$\rightarrow\piz\piz\piz\gamma\rightarrow 7\gamma$ are  
two such states, and can be selected by seeking seven prompt clusters in the
EmC with no charged track in the event.\\
As far as other backgrounds are concerned, only fully neutral channels
are relevant. However, due to high hermeticity of the EmC, the most
relevant of them, the \fikskl$\rightarrow 5\piz \rightarrow 10\gamma$ , may
mimic a 7\phot final state only in a very small fraction of
events. Moreover, the prompt \phot selection rules out all events where \Kl
is not decaying very near the beam pipe. A sample of $5\cdot 10^5$ Monte
Carlo events has been generated for this background, and no event survived
the topological selection cuts. 
The simple, topological selection is common to
both \fietag and \fietapg events and requires:
\begin{itemize}
\item 7 prompt neutral clusters with $21^{\circ}<\theta<159^{\circ}$.
\item No charged tracks.
\item $|E_{tot}-1020 \MeV| < 130 \MeV$
\end{itemize}
\noi where $E_{tot}$ is the sum of the energies of the selected clusters.\\
\noi The efficiency of this cut is $41.3\%$ for \fietag events and
 $41.2\%$ for \fietapg events.
Events passing this cut are further analyzed in both hypotheses of being
\fietag and \fietapg events.
 
To completely rule out any \fikskl background a cut on $E_{\gamma}^{\rm
max}>300 \MeV$ 
can be put for \fietag events with essentially no loss in efficiency.
Monte Carlo simulations, and fit to the \Eta mass peak, show that
background in the \fietag channel is then again at the level of 0.1\%.

No attempt is made to solve the combinatorial for the three \piz's coming
from the \Eta decay, and the only identified photon is the radiative one,
which is, as usual in \fietag, the most energetic photon of the event.

For \fietapg the further analysis is based on a two-steps kinematic fit. 
First, a kinematic with no mass constraints is performed to achieve a
better determination of the photons energies. Then a pairing procedure is
applied in order to obtain the correct identification of the photons. 

The starting point of this procedure is the observation that the most energetic
photon of the event comes always from the \Eta decay. The remaining six
photons are then scanned to check the best pairing giving the correct \Eta
mass. Once the second photon is assigned to the \Eta, the most energetic of
the five remainders is coming from one of the \piz's : Monte Carlo shows
that in this way the
first three photons are correctly assigned in 99.3\% of the cases. 

The remaining 4 photons give rise to twelve possible combinations: a
$\chi^2$-like function is built for each combination to compare the
obtained masses to the expected ones, and among these the best five
combinations are selected : in 96\% of cases among these there is the
correct one. Finally the five best combinations are fitted with a full
constrained 
kinematic fit, where the \piz's , \Eta and \etap invariant masses
constraints are added to the constraints used in the preliminary fit. 
The combination minimizing the final $\chi^2$ is then chosen to be the
correct one, and the $\chi^2$ of this fit is used as a discriminating
variable for background suppression.

The combination found is the correct one in 90\% of the cases, the main
sources of mistakes being the radiative \phot associated incorrectly to a
\piz (5\%) and/or the 4 \phot's of the \piz's  being mismatched (5\%).

The $\chi^2$ corresponding to the best combination is compared to the one
of a kinematic fit performed in the \fietag hypothesis, with only the \Eta
mass as intermediate mass constraint.
Using Monte Carlo events an optimized cut has been chosen in a triangular
shaped region in the plane ${\cal P}(\chi^2_{\etap\gamma})-{\cal
P}(\chi^2_{\eta\gamma})$. Analytically it can be described by the formula:
$$ 
{\cal P}(\chi^2_{\etap\gamma})
 >15\% + 3.4\cdot{\cal P}(\chi^2_{\eta\gamma}) 
$$
This cut alone, although being able to reduce drastically the \fietag
background, is 
not enough to obtain a satisfactory Signal/Background ratio: indeed
$5\times 10^{-4}$
\fietag events still survive the cut against 15.6\% of \fietapg, giving a
S/B ratio of $\approx 0.8$.\\
For this reason a
further selection cut has been introduced.\\
The distribution of the energy of the most energetic photon after the
kinematic fit  is shown in
fig. \ref{enemost} for \fietapg and \fietag fully neutral events. 
%%%%%%%%%%%%%%%%%%%%%%%%%%%%%%%%%%%%%%%%%%%%%%%%%%%%%
\begin{figure}
\begin{center}
\epsfig{file=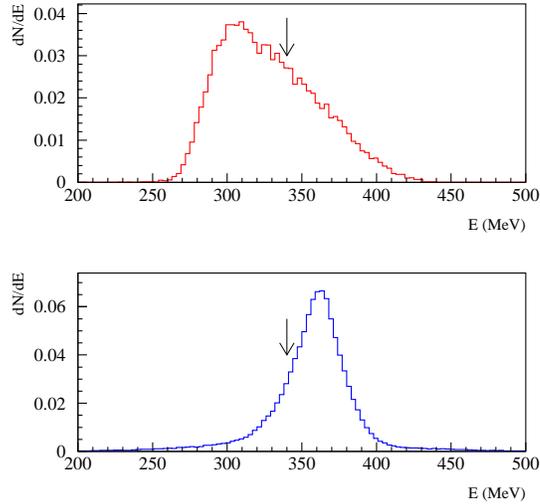,height=.35\textheight}

\caption{{ Monte Carlo energy spectrum of the most energetic
cluster for  
events passing both \fietag and \fietapg event selection, and after the
kinematic fit with no mass constraints.
The plot shows \fietag events (lower plot), \fietapg events (upper plot) and
the position of the selection cut (arrow).}}
\label{enemost}
\end{center}
\end{figure}
%%%%%%%%%%%%%%%%%%%%%%%%%%%%%%%%%%%%%%%%%%%%%%%%%%%%%
A cut on this energy is able to scale down definitively the \fietag
background, even if it causes a somewhat loss in efficiency for the \fietapg
signal. The maximization of the significance lead to the choice of
a cut at $E^{\gamma}_{max} <340 $ MeV.\\
\noi The final selection efficiency for \fietapg events is 
$\varepsilon_{Sel} = 12.7\%$ while selection efficiency for the \fietag
background goes to $\simeq 6.7 \times 10^{-6}$. This results in an
expected S/B ratio $>$ 20 (90\% C.L.) using the PDG'98 value for BR(\fietapg).

Applying the selection criteria described above to the 2.4 \pbinv
statistics of 1999 runs, we select $6^{+3.3}_{-2.2}$ \fietapg$\rightarrow 7
\gamma$ events (with less than one event of  background expected at 90\%
C.L.) and 10938 \fietag$\rightarrow 7 \gamma$ events. This is the first
observation of the decay chain \fietapg$\rightarrow 7\gamma$. 
The distributions of the \fietag control sample compare very favourably with
 the simulations for what the variables on which the cut are set are concerned
 (see plot \ref{plot7gamma}).
%%%%%%%%%%%%%%%%%%%%%%%%%%%%%%%%%%%%%%%%%%%%%%%%%%%%%
\begin{figure}
 \begin{center}
 \begin{tabular}{cc}
\epsfig{file=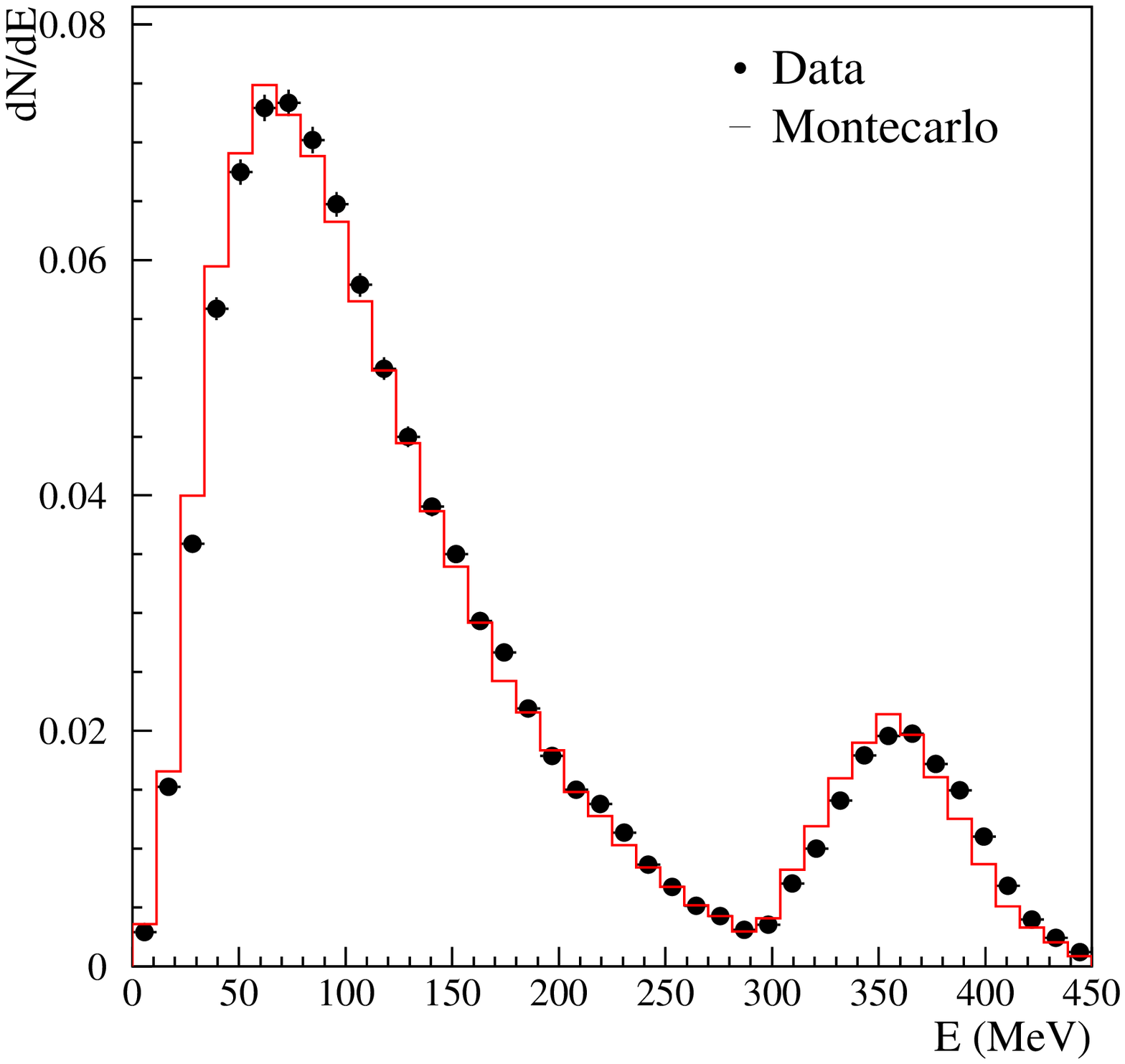,width=48mm}

 &
 \epsfig{file=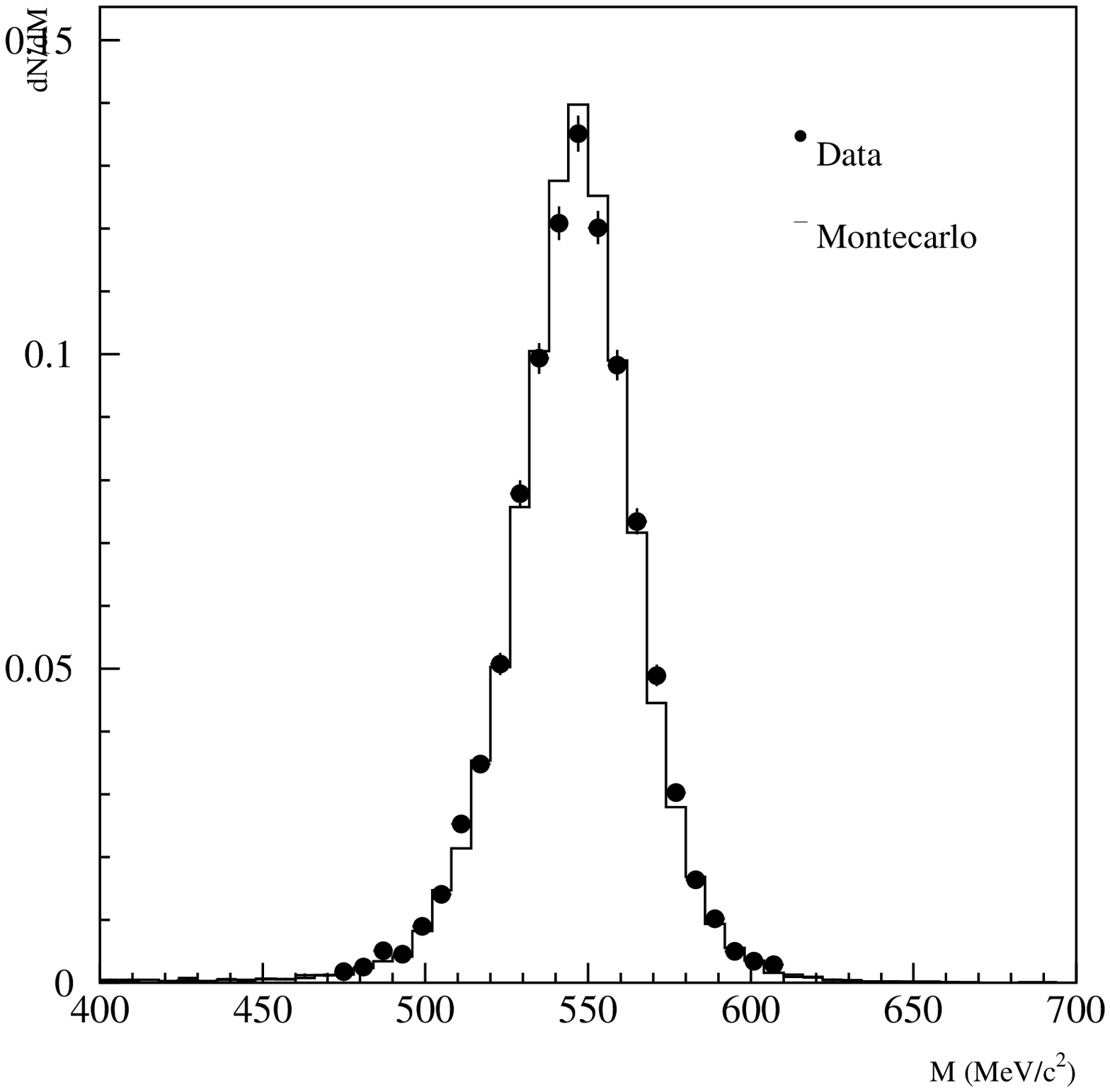,width=48mm}\\
 \end{tabular}
 \end{center}
\caption{{ Monte Carlo (pure \fietag sample) versus data for
7\phot events:
  Cluster energy spectrum (left); \Eta invariant mass (right)}}
 \label{plot7gamma}
\end{figure}
This gives, for $R_{\phi}$:

$$
R_{\phi} =
\frac{N_{\etap\gamma}}{N_{\eta\gamma}}\frac{\varepsilon_{\eta\gamma}}{\varepsilon_{\etap\gamma}}\frac{BR(\etapizpizpiz)BR(\piz\rightarrow\gamma\gamma)}{BR(\etappizpizeta)BR(\etagg)}
$$ 
and, thus: 
$$
R_{\phi} = \left(6.9^{+3.8}_{-2.5} (\rm stat.) \pm 0.9 (\rm
syst.)\right)\cdot 10^{-3}  
$$
where, as already discussed for the \pip\pim\phot\phot\phot final state, 
since most of the systematic effects cancel out in the ratio, the
systematic error is dominated by the efficiency for low energy photons (10\%), 
by the uncertainty on the value of the intermediate branching ratios 
(6\%) and by a 5\% systematic effect evaluated on the efficiency of the 
$\chi^2$ cut due to radiative photon misassignment. 
It is interesting to note that the analysis performed in this channel, 
although being statistically less accurate than the one for 
charged/neutral final state, is fully compatible with that one, and 
constitutes the first measurement of $R_{\phi}$ performed with a decay
chain 
different from the one leading to the \pip\pim\phot\phot\phot final state.

\subsubsection{The  $\eta-\eta'$ mixing angle} 
The importance of the measurement 
of $R_{\phi}$ to extract with great precision the pseudoscalar mixing angle
has been stressed many times during the years \cite{q3}.
If one neglects  $\phi-\omega$ mixing and 
SU(3) breaking effects in the effective Lagrangian, the 
\fietag, \fietapg decays may be described as simple magnetic dipole 
transition, giving for the ratio $R_{\phi}$ the value: 
$$
R_{\phi} = \cot^2 
\varphi_P\left(\frac{p_{\eta'}}{p_{\eta}}\right)^3 
$$
with $\varphi_P = \vartheta_P + \arctan\sqrt 2$. 
In a recent paper by Bramon et al.\cite{BramonEscribano} it has 
been stressed, however, that if one takes into account also $\phi-\omega$ 
mixing angle $\varphi_V = +3.4^{\circ}$, and accounts for SU(3) breaking via a
term proportional to 
$\frac{m_s}{\bar{m}}\simeq 1.45$ the formula above gets a correction 
factor: 
$$
R_{\phi} = \cot^2
\varphi_P\left(1-\frac{m_s}{\bar{m}}\frac{\tan\varphi_V}{\sin
2\varphi_P}\right)^2 \left(\frac{p_{\eta'}}{p_{\eta}}\right)^3  
$$
The formula above has been used, together with the measured value of
$R_{\phi}$ to extract a measurement for \tp, giving: 

$$
\vartheta_P =
\left(-18.9^{\circ}{}^{+3.6^{\circ}}_{-2.8^{\circ}}(\mbox{stat.}) \pm
0.6^{\circ}  (\mbox{syst.})\right) 
$$

\section{$\phi\rightarrow\pi^{+}\pi^{-}\pi^{0}$}

$\pi^+\pi^-\pi^0$ events are selected requiring a prompt vertex with two
opposite sign 
tracks and two prompt photons in the calorimeter. From the two tracks, the
direction of the missing momentum can be evaluated and associated with the
$\pi^0$ direction. The opening angle between the two 
photons in the $\pi^0$ rest frame is required to be larger than
$170^o$. Furthermore in order to remove a residual background mainly due to
$e^+e^-\gamma\gamma$ an opening angle between the two tracks less than
$170^o$ is also required. 

The final sample (330000 events) has been analyzed by means of the Dalitz-plot
method.

%The data can be analyzed by means of the Dalitz-plot
%method. Fig.\ref{Dalitz} shows the Dalitz-plot obtained using about 330000
%events selected according to the previously defined cuts.
%The events are all well within the kinematical contour, and the shape of
%the bi-dimensional distribution is essentially determined by the dominance
%of the $\rho\pi$ contribution that results in the small amount of events at
%the corners of the plot.

%\begin{figure}
%\begin{center}
%\epsfig{file=newdal_3.eps,height=.35\textheight}
%\caption{Dalitz-plot distribution of the selected $\pi^+\pi^-\pi^0$ events.
%The energy difference between charged pions (horizontal axis) is plotted vs.
%the kinetic energy of the $\pi^0$ (vertical axis). Also shown is the
%kinematical contour of the allowed region.}
%\label{Dalitz}
%\end{center}
%\end{figure}
 
%A significant dependence of the selection efficiency on the position on the
%$Dalitz-plot has been found using a complete simulation of the KLOE
%detector. A decrease in efficiency is found mostly in the two upper corners
%(see Fig.\ref{Dalitz}) both corresponding to events with at least one
%low-momentum track. A correction function has been found after the
%simulation of $10^6$  $\pi^+\pi^-\pi^0$ final states uniformly generated on
%all the Dalitz-plot. 

The Dalitz-plot binned in $8\times8~MeV$ squares and corrected for the
efficiency is fitted to a model of  $\pi^+\pi^-\pi^0$ production including
the following terms: 
\begin{enumerate}
\item{{\bf $A_{\rho\pi}$} is the $\rho\pi$ amplitude given by the sum of
the three $\rho$ charged states, each described by a Gounaris-Sakurai
parametrization. Free parameters are the $\rho$ masses and the width;} 
\item{{\bf $A_{direct}$} is the direct term contribution given by a complex
number that is two free parameters, namely a modulus and a phase. The
modulus is normalized in such a way that a value equal to 1 corresponds to
a direct term equal to the $\rho\pi$ term.}
\item{{\bf $A_{\omega\pi}$} is the $\omega\pi$ term, where mass and width
of the $\omega$ are fixed to the PDG values, and only a
complex amplitude that is a modulus and a phase is let free.} 
\end{enumerate}
 
The fitting function is then given by ($X$ and $Y$ are two Dalitz variables):
$$f(X,Y)=|\vec{ p}^{~+}\times \vec{p}^{~-}|^2\cdot| {\bf A_{\rho\pi}}+{\bf
A_{direct}}+{\bf A_{\omega\pi}}|^2$$ 
where the square of the vector product in front takes into account the
vector nature of the decaying particle. 

\begin{table}
  \begin{center}
    \begin{tabular}{|c|c|c|}
\hline
Parameter & Fit result & PDG result \\ \hline
 $M(\rho_0)$ (MeV) & $776.1\pm 1.0$ & $776.0\pm 0.9^{*}$ \\
 $\Delta M$ (MeV) & $-0.5\pm 0.7$ & $0.1\pm 0.9$ \\
 $\Gamma(\rho)$ (MeV) & $145.6\pm 2.2$ & $150.9\pm 2.0$\\
 A(direct term)/A($\rho\pi$) & $0.10\pm 0.01$ & $-0.15\div 0.11$\\
 fase(direct term)-fase($\rho\pi$) & $(114\pm 12)^o$ & \\
\hline
    \end{tabular}
  \end{center}
\caption{Results of the fit to the Dalitz plot compared to the PDG values.}
\label{Tableppp}
\end{table}

In Table.\ref{Tableppp} the results of the fit are shown and compared with
PDG values. 
Two observations can be done. 

First we observe a sizeable direct term (about $10~\%$ of the
$\rho\pi$ term) with a phase respect to $\rho\pi$ loosely close to
$90^o$. We remark that this is the first observation of this decay. 

Second we find values of the $\rho$ line-shape parameters that are in
agreement with PDG numbers. The mass is in agreement with the one obtained
in $e^+e^-$ experiments. Furthermore the mass difference between the
neutral and the charged $\rho$s is compatible with 0, so that no isospin
violations are observed. The latter 
results improves the PDG values.

\pagebreak

\end{document}